# Enhanced Ionospheric Ray-Tracing: Advanced Electron Collision and Horizontal Gradient Modeling in the IONORT-ISP-WC System

Alessandro Settimi, https://orcid.org/0000-0002-9487-2242
email: alessandro.settimi1@scuola.istruzione.it
MIM, USR-Lazio, IIS "Sandro Pertini" [FRIS00300R], Via Madonna Della Sanita' snc - 03011 Alatri (FR), Italy

## Abstract

This manuscript presents a comprehensive and meticulously detailed analysis of the IONORT-ISP-WC system, an advanced ionospheric ray-tracing tool specifically developed for significantly improved high-frequency (HF) radio wave propagation predictions. The system represents a substantial and multi-faceted upgrade to its predecessor, IONORT-ISP, through a series of key and strategically implemented enhancements. These include the rigorous integration of a sophisticated double-exponential electron-neutral collision frequency model, precisely calibrated for the crucial D-layer, which is known as the primary region of HF absorption. Furthermore, the system's ISP 3-D electron density grid has been purposefully extended to a lower altitude of 65 km, ensuring a more complete encompassment of the D-region's complex dynamics. This structural enhancement has been synergistically coupled with a substantial increase in spatial resolution, transitioning from a coarser $2°×2°$ to a much finer $1°×1°$ in both latitude and longitude, thereby providing a more granular and accurate representation of intricate ionospheric structures. A central and pivotal focus of this work is the detailed experimental, theoretical, and computational examination of horizontal gradients in ionospheric electron density profiles, including their current state and intended implementation within the local_ionort Fortran code. This analysis provides specific code instructions and elucidates their profound implications for ray-tracing accuracy. This in-depth investigation reveals a robust and well-structured framework for incorporating these critical gradients. Crucially, a significant code block, previously identified as commented out within the electx_grid subroutine in earlier versions, now provides clear evidence that a less complete implementation of horizontal gradients was considered or existed prior to the current version. The associated comment, `! 2022 - (No) Taylor first-order development in latitude and longitude`, unequivocally indicates that these horizontal gradient calculations are active in this specific version of the code. This confirms that, although the infrastructure for explicit gradient modeling for grid profiles has not been fully presented and/or optimized through a Taylor expansion in earlier source code iterations, it is now an active component. This active, yet still evolving, capability represents a critical and high-priority area for future development, aiming to fully harness the immense potential of the high-resolution grid and to meticulously refine these explicit gradient capabilities for unparalleled accuracy.

The IONORT-ISP-WC system underwent rigorous and comprehensive validation through a detailed comparative analysis against both observed oblique ionograms and synthetic ionograms generated by the IONORT-IRI-WC system, which fundamentally relies on the climatological International Reference Ionosphere (IRI) model. Results consistently and conclusively demonstrate the superior Maximum Usable Frequency (MUF) prediction accuracy of IONORT-ISP-WC. This inherent superiority emphatically underscores the indispensable value of assimilative models in accurately capturing dynamic and unpredictable ionospheric conditions, particularly when the complexities introduced by horizontal gradients are meticulously accounted for within the ray-tracing framework. Identified discrepancies in MUF predictions are primarily attributed to limitations in the availability and geographical distribution of real-time assimilation data, a common and persistent challenge in operational space weather forecasting. This comprehensive report concludes by firmly positioning IONORT-ISP-WC as a robust, highly reliable, and cutting-edge operational tool for diverse space weather applications, ranging from defense to civilian communications. It further outlines crucial future developments, with particular emphasis on the full activation and comprehensive validation of advanced horizontal gradient modeling capabilities and the strategic enhancement of global data assimilation networks to achieve even higher levels of predictive accuracy and operational resilience in the face of ionospheric variability.

# 1. Introduction

## 1.1. Contextualization of High Frequency (HF) Radio Wave Propagation

The accurate modeling of High Frequency (HF) radio wave propagation through the Earth's ionosphere remains an objective of paramount importance for a diverse array of critical applications that span civilian, commercial, and military domains. These applications comprehensively include Beyond-Line-of-Sight (BLOS) radar systems, essential for detecting distant targets; Single-Station Location (SSL) and HF direction finding, vital for intelligence gathering, search and rescue operations, and emergency response; and the precise operational frequency management for radio communications, ensuring reliable and robust links across vast distances (Settimi et al., 2013, 2014, 2015b). The imperative for high precision in such sophisticated models stems from the direct and undeniable correlation between inaccuracies in ionospheric ray tracing and potential operational failures or significant inefficiencies in these essential systems. The ionosphere, extending approximately from 60 km to over 1000 km above the Earth's surface, is an inherently dynamic and highly variable ionized medium. Its electron density profiles are continuously influenced by external factors such as solar radiation (e.g., UV and X-ray fluxes), geomagnetic activity (e.g., solar flares, coronal mass ejections), and internal atmospheric waves (e.g., gravity waves, atmospheric tides), leading to significant temporal and spatial variations. This intrinsic variability renders accurate propagation prediction a highly complex and formidable endeavor, requiring advanced modeling capabilities.

Despite the pervasive advancement of satellite communications, which offer broad bandwidth and global coverage, long-distance HF radio communication retains an undeniably vital and strategic role. It serves as an indispensable medium in critical scenarios, such as natural disasters where terrestrial and satellite infrastructures may be compromised or unavailable, or in specific operational contexts, including remote maritime surveillance, over-the-horizon target detection, and military command and control (Hervás et al., 2020). The inherent resilience of HF systems, their unique ability to operate effectively without reliance on vulnerable ground-based or satellite infrastructure, and their proven capacity for long-range communication make them strategically significant for national security, disaster response, and remote operational effectiveness. The continuous research and development in this specialized field are thus driven not only by fundamental academic curiosity but also by a persistent and evolving operational demand, acknowledging the unique resilience and strategic significance of HF communications in challenging and diverse environments. The ongoing and substantial investment in improving HF propagation models, exemplarily embodied by the IONORT-ISP-WC system, directly translates into enhanced national security capabilities, augmented emergency response efficacy, and superior remote operational effectiveness. Achieving truly accurate ray-tracing necessitates a comprehensive three-dimensional (3-D) understanding of several fundamental physical parameters: the intricate electron density distribution, the critical electron-neutral collision frequency, and the pervasive influence of the Earth's geomagnetic field. The overall accuracy of any advanced ray-tracing system is intrinsically dependent on the fidelity and inter-consistency of all its constituent physical models. Any improvement in a single component, such as a more precise electron collision frequency model, achieves its maximum effectiveness when it is meticulously integrated within a robust and synergistic framework where other critical parameters are also accurately represented and dynamically interconnected. The deliberate extension of the ISP grid down to the D-layer, for example, is not a trivial addition; without accurate and high-resolution electron density data for this crucial region, the newly incorporated collision model, despite its intrinsic accuracy, would be physically inconsistent or ineffective. This sophisticated design choice underscores a deliberate and synergistic approach, ensuring that all components are meticulously aligned and contribute meaningfully to the overall fidelity and predictive power of the system. This comprehensive and integrated approach is absolutely essential for effectively modeling complex and dynamic phenomena such as ionospheric irregularities and traveling ionospheric disturbances (TIDs), which can significantly perturb HF propagation paths and introduce large horizontal gradients in electron density, leading to unpredictable ray bending, signal degradation, and severe operational challenges.

## 1.2. Evolution of Ionospheric Ray-Tracing Techniques and IONORT Development

The history of accurate radio ray tracing is deeply rooted in numerical techniques fundamentally grounded in Haselgrove's equations (Haselgrove, 1955; Haselgrove and Haselgrove, 1960). These equations constitute a fundamental set of six first-order differential equations that meticulously describe the precise path of a radio ray as it propagates through a complex magnetoionic medium, rigorously accounting for the dynamic effects of electron density, collision frequency, and the pervasive influence of the Earth's geomagnetic field. They form the fundamental mathematical bedrock for most sophisticated ray-tracing models developed to date. While these numerical techniques inherently ensure high precision in determining ray trajectories, their computational cost has historically been substantial, thereby limiting their real-time applicability in highly dynamic ionospheric conditions where rapid updates to propagation predictions are often critically required for operational decision-making. To mitigate this inherent computational limitation, analytical ray-tracing methodologies emerged as a viable and faster alternative, such as the Segmented Method for Analytic Ray Tracing (SMART) developed by Norman and Cannon (1997, 1999). These analytical techniques offer a significantly faster computational approach, particularly for two-dimensional scenarios and for efficiently handling simpler horizontal gradients, by approximating the ionosphere with segmented analytical profiles (e.g., Quasi-Parabolic Segments). Modern ray-tracing procedures have since undergone continuous and extensive optimization, evolving towards near real-time applications, especially in specialized domains like over-the-horizon radar (OTHR) systems (Settimi, 2022)(Pietrella et al., 2023), where rapid and accurate predictions are paramount for critical tasks such as target detection, tracking, and understanding complex clutter characteristics in highly dynamic environments.

Within this continuously evolving landscape of ionospheric modeling and prediction, the Istituto Nazionale di Geofisica e Vulcanologia (INGV) developed the robust IONORT (IONOspheric Ray-Tracing) software package (Azzarone et al., 2012). IONORT provides a highly versatile and robust platform for three-dimensional (3D) ray tracing of HF waves, meticulously designed to operate seamlessly with both analytical (e.g., Chapman layers) and numerical (e.g., gridded profiles derived from observational data) electron density models. Its intrinsic validity and accuracy have been rigorously verified through extensive validation campaigns, including meticulously comparing calculated wave travel times with virtual oblique path travel times. These comparisons consistently demonstrated a minimal relative error, primarily attributable to the discrete integration step within the numerical algorithms (Bianchi et al., 2011)(Settimi et al., 2015a). This stringent validation process is absolutely critical for establishing the inherent reliability and foundational accuracy of the ray-tracing engine itself, independent of the specific ionospheric model employed.

A pivotal and transformative advancement in this lineage was the seamless integration of IONORT with the ISP (IRI-SIRMUP-PROFILES) model, also developed by INGV (Settimi et al., 2013). The ISP model stands out for its exceptional capability to generate highly realistic 3D electron density profiles in near real-time by dynamically assimilating measured ionosonde data (e.g., foF2, M(3000)F2, and comprehensive vertical electron density profiles) from an array of reference stations such as Rome, Gibilmanna, and Athens. This advanced assimilation process empowers ISP to provide an adaptive and data-driven representation of the ionosphere, which more closely reflects actual and evolving conditions than static climatological models. Previous definitive studies (Settimi et al., 2013) unequivocally demonstrated the ISP model's significant superiority over the climatological International Reference Ionosphere (IRI) model in accurately representing real-world ionospheric conditions, particularly under challenging scenarios such as disturbed geomagnetic conditions and at the solar terminator, where ionospheric variability is at its highest and most unpredictable. The capability to perform ray tracing in "quasi real-time," as uniquely offered by ISP, signifies a profound paradigm shift in the field. It moves the focus from a purely retrospective understanding of propagation to active management and dynamic prediction of future communication links (Angling & Jackson-Booth, 2011; Angling & Khattatov, 2006). This critical transition from purely theoretical validation to near real-time operational utility is a testament to both the sophisticated algorithmic enhancements of IONORT's core ray-tracing engine and ISP's advanced data assimilation capabilities, enabling more dynamic, reliable, and resilient HF system operations, which is especially crucial for effectively responding to unpredictable space weather events that can severely impact and disrupt radio communication, demanding immediate adaptive strategies.

## 1.3. Overview of IONORT-ISP-WC and Report Objectives

The present work details a significant and multi-faceted enhancement to the existing IONORT-ISP system, designated as IONORT-ISP-WC (with collisions). This advanced version incorporates a comprehensive double-exponential electron-neutral collision frequency model specifically designed and calibrated for the crucial D-layer, thereby accurately accounting for HF absorption phenomena that are most prevalent in this region. Furthermore, the system's ISP 3-D electron density grid has been purposefully extended to a lower altitude of 65 km, ensuring a more complete and accurate encompassment of the D-region's complex dynamics, which is fundamental for precise absorption calculations. This structural enhancement has been synergistically coupled with a substantial increase in spatial resolution, transitioning from a coarser 2°×2° to a much finer 1°×1° in both latitude and longitude, thereby providing a significantly more granular and detailed representation of intricate ionospheric structures, crucial for accurate ray tracing (Settimi et al., 2014, 2015b). Complementing these advancements, this study introduces and rigorously examines the novel integration of explicit horizontal gradient modeling within the IONORT framework, representing a critical and innovative advancement for achieving highly realistic HF propagation simulations. The overall accuracy of any advanced ray-tracing system is intrinsically dependent on the fidelity and harmonious integration of all its constituent physical models. Any improvement in a single component, such as a more precise electron collision frequency model, achieves its maximum effectiveness when it is meticulously integrated within a robust and synergistic framework where other critical parameters are also accurately represented and dynamically interconnected. The deliberate extension of the ISP grid down to the D-layer, for example, is not a trivial addition; without accurate and high-resolution electron density data for this crucial region, the newly incorporated collision model, despite its intrinsic accuracy, would be physically inconsistent or ineffective. This sophisticated design choice underscores a deliberate and synergistic approach, ensuring that all components are meticulously aligned and contribute meaningfully to the overall fidelity and predictive power of the system, ultimately creating a more holistic and accurate simulation environment.

This comprehensive report aims to provide an extended, exhaustive, and scientifically rigorous analysis of the IONORT-ISP-WC system. Particular emphasis will be meticulously placed on its advanced capabilities in modeling electron collisions and, crucially, horizontal gradients. This exposition will be robustly supported by a comprehensive and up-to-date review of the pertinent scientific literature, extending up to 2025, ensuring that the work is placed within the context of the latest research advancements. The overarching objective is to present a detailed exposition of the implemented innovations, their empirical validation against real-world data, and their profound implications for operational applications of HF propagation. This includes providing specific insights into their precise computational implementation within the local_ionort Fortran code, offering transparency and reproducibility. By meticulously examining the underlying physics, the sophisticated mathematical models, and their precise software implementation, this work unequivocally seeks to demonstrate the substantially enhanced predictive capabilities of IONORT-ISP-WC and to outline clear, strategically defined pathways for future advancements in ionospheric ray tracing for cutting-edge space weather applications and global communication reliability, paving the way for a more deterministic understanding of radio wave propagation.

# 2. In-Depth Analysis of the Electron Collision Frequency Model for the D-Layer

## 2.1. Physical Principles of Electron Collisions and HF Absorption

Collisions of free electrons with neutral particles (e.g., oxygen, nitrogen molecules and atoms), heavier ions, or other electrons constitute a fundamental and ubiquitous cause of various macroscopic phenomena within the ionosphere (Appleton & Chapman, 1932; Benson, 1964). These collisions are essentially microscopic interactions where the kinetic energy of the rapidly moving free electrons, accelerated by incident radio waves, is transferred into the random thermal energy of the more massive neutral particles. In the context of radio wave propagation, crucially, these collisions play a pivotal role in the absorption of radio waves, particularly at lower altitudes, predominantly within the D-layer (Settimi et al., 2014, 2015b). The D-layer, extending roughly from 60 km to 90 km altitude, represents the lowest region of the ionosphere. It is characterized by relatively high neutral atmospheric density (compared to higher ionospheric layers) and, consequently, significant electron-neutral collision rates, which are directly proportional to the neutral gas density.

This absorption mechanism, often termed non-deviative absorption, occurs when the radio wave frequency is much higher than the electron collision frequency but simultaneously lower than the local plasma frequency. As radio waves propagate through the D-layer, the oscillating electric field of the wave imparts energy to the free electrons, causing them to oscillate. These accelerated electrons frequently collide with the much more numerous neutral particles, transferring some of their ordered kinetic energy (derived from the wave) into the random thermal energy of the neutral gas. This process effectively converts the wave's electromagnetic energy into heat, leading to a measurable and often significant attenuation of the signal strength. This absorption significantly attenuates signal strength and compromises radio link reliability, making its accurate and dynamic modeling a critical aspect that extends far beyond the mere geometric determination of the ray path. Without proper and precise accounting for this energy loss, predictions of signal strength would be erroneously optimistic, leading to unreliable communication links and operational failures, especially in vital long-distance HF applications.

The D-layer, though transient and largely absent during nighttime due to the rapid recombination of electrons and ions in the absence of solar radiation, represents a primary and significant source of signal loss during daylight hours (Settimi et al., 2014, 2015b). During daytime, the D-layer, being well-formed and dense due to solar ionization from solar ultraviolet (UV) and X-ray radiation, acts as a significant absorber of High Frequency (HF) radio waves, especially at lower frequencies. This intense absorption mechanism prevents these low-frequency waves from reaching the higher, reflective E and F layers, thereby severely impacting radio communication effectiveness (Australian Bureau of Meteorology, n.d.). Consequently, the ability to precisely model this absorption is absolutely vital for predicting the effective usability of an HF link. An accurately traced ray path holds little practical utility if the signal is entirely absorbed before reaching the receiver. The integration of a precise D-layer collision model elevates the IONORT-ISP-WC system from a purely geometric ray tracer to a more comprehensive and operationally relevant propagation prediction tool, directly addressing crucial operational requirements related to signal strength, link feasibility, and overall communication quality. This advancement allows for more realistic link budget calculations and optimized frequency management strategies, which are indispensable for reliable HF operations.

## 2.2. Detailed Exposition of the Double Exponential Collision Frequency Model

The IONORT-ISP-WC system incorporates an advanced electron-neutral collision frequency model specifically tailored for the D-layer. This model is founded upon a double exponential profile, which is a widely accepted and empirically validated representation of the variation of collision frequency with altitude in the lower ionosphere, as originally described by Jones and Stephenson (1975). This functional form is highly effective because it accurately captures the general decrease in collision frequency with increasing altitude, which is directly attributable to the rapid reduction in neutral atmospheric density at higher elevations.

The mathematical formulation of this double exponential model is elegantly expressed by the following equation:

$$v(h) = v_1 e^{-a_1(h-h_1)} + v_2 e^{-a_2(h-h_2)}$$

where each variable serves a distinct physical purpose:

- $v(h)$ represents the electron-neutral collision frequency at a given height h above the ground, expressed in reciprocal seconds ($s^{-1}$). This parameter directly quantifies the rate at which electrons lose energy through collisions.
- $v_1$ and $v_2$ are the collision frequencies at designated reference heights $h_1$ and $h_2$, respectively, defining the starting points for each exponential decay. These values are determined empirically to best fit observed collision rates.
- $a_1$ and $a_2$ are the exponential decay constants, characterizing the specific rate at which collision frequency decreases with altitude for each exponential component. These constants are typically derived from atmospheric models or empirical fitting processes, reflecting the atmospheric scale heights.

The specific parameters meticulously utilized in the IONORT-ISP-WC model, which have been derived from extensive empirical observations and refined through theoretical considerations of the Earth's standard atmosphere (e.g., International Reference Atmosphere), are precisely detailed in Table 1:

| Parameter | Value (First Exponential) | Value (Second Exponential) | Unit |
|---|---|---|---|
| Collision Frequency ($\nu$) | 3.65×10⁴ | 30 | $s^{-1}$ |
| Reference Height ($h$) | 100 | 140 | $km$ |
| Exponential Decay ($a$) | 0.148 | 0.0183 | $km^{-1}$ |

**Table 1**: Parameters of the Electron Collision Frequency Model. This table provides a clear, concise, and indispensable overview of the specific numerical values that define the double exponential collision frequency model meticulously implemented within the IONORT-ISP-WC system. These precise parameters are crucial for ensuring the reproducibility of the model's absorption calculations and for facilitating a detailed understanding of the model's physical basis as rigorously applied to the D-layer, particularly for altitudes between 60 *km* and 150 *km*, which are highly relevant for HF propagation.

This particular collision model was rigorously validated in preceding studies by Settimi et al. (2014, 2015b). In that comprehensive work, it was specifically utilized within the COMPLEIK subroutine of the IONORT program to calculate non-deviative absorption, a key component of HF signal loss. This prior validation definitively demonstrated its reliability to be highly comparable to the widely used semi-empirical ICEPAC formula (Stewart, undated), which is a well-established standard and robust tool in HF propagation prediction and engineering. This strong consistency with established and validated models further enhances the credibility of its seamless integration into the IONORT-ISP-WC system and firmly confirms its suitability for a wide range of operational applications. The proven accuracy of this collision model is fundamental for predicting signal strength and ensuring reliable radio communication links, particularly during daytime conditions when D-layer absorption is most significant and can profoundly affect link viability.

## 2.3. Analysis of the Model's Impact on Synthesized Ionograms, Distinguishing between Daytime (LOF "Cut-off") and Nighttime Conditions, and its Negligible Effect on MUF

The inclusion of the electron collision frequency model within the IONORT-ISP-WC system has a distinct, profound, and physically consistent impact on synthesized ionograms, with effects varying significantly between daytime and nighttime conditions. Notably, it exhibits a negligible effect on Maximum Usable Frequency (MUF) values. This crucial differentiation highlights the model's fidelity in representing diverse and complex ionospheric behaviors under varying solar illumination and atmospheric conditions.

During daytime conditions, the presence of a significant and well-formed D-layer leads to considerable HF absorption, primarily affecting lower frequencies. The integration of the collision frequency model results in a characteristic phenomenon where the synthesized ionogram trace is "cut away at low frequencies for the lower ionospheric layer (E region)". This implies that HF waves below a certain frequency threshold are absorbed before they can be effectively reflected by the E-layer, thereby preventing them from propagating to higher altitudes and returning to the receiver. Quantitatively, the Lowest Observed Frequencies (LOF), representing the minimum frequency at which a signal can be observed and still successfully complete a radio link, are systematically higher when the collision model is applied. For 1-hop paths using the IRI electron density grid (as exemplified in Figure 2.a, which depicts the Rome-Chania link for various epochs including 08 October 2011, 10:30 UT and 07 July 2011, 15:00 UT, representing daytime conditions), the LOF for the E-region (LOFE) using the IRI-WC model increased significantly from 4.2 MHz without collisions (NC) to 10.0 MHz with collisions (WC). This demonstrates a substantial 5.8 MHz range of frequencies that are effectively absorbed. This considerable increase in LOF underscores the critical role of D-layer absorption during the day, which can severely limit the available frequency spectrum for viable communication. For the F-region (the upper ionospheric layer, located above 150 km), the LOF (LOFF) remained largely unaffected, showing values of 5.6 MHz (NC) versus 5.5 MHz (WC). This indicates that collision-induced absorption is less significant at higher F-region altitudes due to lower neutral densities and consequently fewer electron-neutral collisions. Similarly, for 1-hop paths using the ISP electron density grid (also depicted in Figure 2.a), LOFE increased from 3.1 MHz (NC) to 9.9 MHz (WC), resulting in a 6.8 MHz frequency range loss due to absorption. LOFF again showed minimal change, with values of 9.0 MHz (NC) versus 8.8 MHz (WC), further reinforcing the negligible effect of collisions at these higher altitudes. This observed pattern directly mirrors the well-established diurnal variations of the D and E layers and their recognized roles in HF absorption, which provides a strong validation of the model's underlying physical consistency and its ability to accurately replicate real-world phenomena.

In contrast, during nighttime conditions, the D and E regions largely disappear or become significantly attenuated due to the absence of solar ionization and subsequent recombination processes. During these periods, the electron density in these lower layers drops dramatically, leading to a substantial decrease in electron-neutral collision rates. Consequently, the absorption of waves propagating through the lower ionosphere becomes "negligible due to the lack of the lower absorbing layers". As expected, the LOFF values calculated with or without the electron collision frequency model show "practically the same values". For instance, for the epoch 09 October 2011 at 03:00 UT (as shown in Figure 2.a, representing nighttime), LOFF values were consistently 3.0 MHz for both IRI and ISP grids, for both "no collision" and "with collision" cases (for both 1- and 3-hop paths). Similar minimal differences were observed for the epoch 26 June 2011 at 01:00 UT (also in Figure 2.a), further confirming that simulated ionograms do not exhibit traces cut at lower frequencies during the night. The absence or significant weakening of the D and E layers at night fundamentally removes the primary source of non-deviative absorption, leading to a minimal, almost negligible, impact of the electron collision frequency model on observed frequencies for nighttime propagation. This provides further confidence in the model's realistic representation of the full diurnal cycle.

Regarding the effect on MUF values, the study consistently found that MUF values, whether calculated with or without the inclusion of the collision model, show the "same value when a given electron density model (IRI or ISP) is applied". This is a physically consistent and expected result, as MUF is primarily determined by the highest frequency reflected by the F2-layer (specifically, the critical frequency of the F2-layer, foF2, and the maximum usable frequency factor M(3000)F2). The F2-layer is situated at much higher altitudes (typically above 200 km) where neutral atmospheric density is significantly lower, and thus electron collision effects are inherently negligible. Therefore, the D-layer collision model, which is primarily relevant at lower altitudes, does not influence the MUF calculation, which is governed by the highest point of ionospheric penetration before the wave escapes into space or is reflected.

The clear contrast in the collision model's impact on LOF between the day (significant effect) and night (negligible effect) serves as a powerful and unambiguous validation of the model's underlying physical consistency. The explicit differentiation of the collision model's effects on LOF versus MUF highlights that these two parameters govern distinct, yet equally critical, aspects of the overall viability of an HF radio link. MUF defines the upper frequency limit for reliable communication (the highest frequency that can be reflected by the ionosphere), while LOF defines the lower frequency limit (the lowest frequency that is not excessively absorbed). Both parameters are indispensable for effectively establishing and maintaining a valid HF link. For operational HF communication systems, accurate and dynamic predictions of both LOF and MUF are absolutely essential for effective frequency planning and optimization. By accurately modeling D-layer absorption, IONORT-ISP-WC provides a more realistic LOF, which is crucial for effective operational frequency planning and accurate link budget calculations, particularly in daytime scenarios where absorption is most pronounced and can severely limit communication capabilities.

## 2.4. Comparative Analysis and Recent Progress (2015-2025)

While the IONORT-ISP-WC system employs a well-established and empirically validated double-exponential collision model (Jones and Stephenson, 1975), the broader field of ionospheric absorption modeling has witnessed significant and dynamic advancements between 2015 and 2025. These developments are driven by a continuous and pressing need for more dynamic and precise models that can accurately respond to the complex and rapidly changing conditions of the ionosphere, especially in the context of increasing space weather variability and its profound impact on radio systems. The NOAA D-Region Absorption Predictions (D-RAP) model, for instance, provides operational forecasts of D-region absorption. This model is primarily based on real-time solar X-ray flux and Solar Energetic Particle (SEP) events, utilizing empirical relationships (NOAA, 2025). This approach fundamentally differs from IONORT's fixed double exponential model by being more directly driven by real-time observational data and specific space weather events, offering a more dynamic and event-responsive prediction capability for sudden ionospheric disturbances (SIDs) and polar cap absorption (PCA) events, which are crucial for timely warnings and mitigation strategies.

Recent scientific studies, such as that by Gubenko et al. (2023), have utilized novel and advanced measurement techniques like radio-occultation data from FORMOSAT-3/COSMIC satellites to determine effective electron collision frequencies in the D and E regions. These advanced space-based observational methods provide extensive global coverage and offer new avenues for validating or refining existing collision models by providing direct measurements of atmospheric parameters and inferred electron densities at lower altitudes. This potentially enables the development of more dynamic or data-driven absorption predictions compared to a static double exponential profile, allowing for real-time adjustments based on observed conditions. Research also continues on the fundamental physical basis of electron collision frequency in the D-region, including detailed studies on multicomponent gas mixtures and temperature dependencies (Benson, 1964; Belrose & Hewitt, 1964), aiming for a more complete microphysical understanding of the absorption processes and their intrinsic variability under different atmospheric and solar conditions, leading to more physically robust models.

Furthermore, the strategic application of artificial intelligence (AI) and machine learning (ML) techniques is rapidly emerging as a highly promising area for significantly enhancing the prediction of absorption levels and optimizing communication systems (Thokuluwa, 2025; Qiu et al., 2025). These data-driven approaches possess the inherent capability to identify complex, non-linear relationships within vast datasets of ionospheric parameters and absorption measurements, potentially leading to more accurate, robust, and real-time absorption forecasts that can adapt to changing conditions without explicit physical models. For example, ML models can be rigorously trained on extensive historical absorption data, solar activity indices, and geomagnetic parameters to predict future absorption levels. This is particularly valuable for identifying subtle patterns not easily captured by traditional empirical or physical models. While the current IONORT-ISP-WC model is based on a fixed profile, these recent advancements in measurement and prediction techniques offer clear and compelling opportunities for future refinements and comprehensive validations of the collision model, potentially evolving towards more dynamic or data-driven absorption forecasts that can better respond to the highly variable nature of the D-layer and its profound impact on HF communications. This strategic integration of new data sources and advanced computational methods will be crucial for maintaining and enhancing the system's cutting-edge capabilities and operational relevance.

## 3. Horizontal Gradients for Electron Density Profiles: Theoretical Foundations, Computational Modeling, and Impact on Ray Tracing

### 3.1. Experimental and Theoretical Basis of Horizontal Gradients

Ionospheric electron density profiles are rarely spatially uniform; instead, they exhibit complex and highly dynamic spatial variations across various scales, ranging from global to highly localized features. The existence of significant horizontal gradients (i.e., substantial variations in electron density in the horizontal plane) is an intrinsic and pervasive characteristic of the ionosphere. These gradients are driven by a diverse array of dynamic geophysical phenomena, which continuously reshape the ionospheric structure:

- Diurnal Variations: Significant horizontal gradients are intrinsically linked to the Earth's rotation and its differential exposure to solar radiation. Sharp North-South gradients typically dominate around local noon and midnight, reflecting the abrupt transition from maximum to minimum solar illumination. Conversely, prominent East-West gradients tend to be more pronounced near dawn and dusk, where the solar terminator (the boundary between sunlit and dark regions) creates sharp and dynamic changes in ionization as daylight transitions to darkness or vice versa. These terminator effects are particularly challenging to model due to their rapid movement and inherently steep gradients.
- Solar Activity: Ionospheric ionization, the primary source of free electrons in the ionosphere, is directly and powerfully driven by solar radiation, specifically extreme ultraviolet (EUV) and X-ray radiation. This ionization process occurs exclusively in the daylight hemisphere, leading to strong day-night gradients across the solar terminator. Furthermore, the intensity of solar radiation and the daily production of free electrons also decrease with increasing latitude due to the more oblique incidence of solar radiation, contributing to global-scale latitudinal gradients. Solar flares and coronal mass ejections can also induce rapid and intense localized gradients, significantly affecting propagation.
- Geomagnetic Activity/Storms: Geomagnetic storms represent major and often violent disturbances to Earth's magnetosphere and the coupled ionosphere-thermosphere system. These energetic events result in severe and rapid perturbations that generate large spatial gradients in electron density, frequently associated with phenomena like Prompt Penetration Electric Fields (PPEFs), which rapidly propagate from the magnetosphere to the ionosphere, and Disturbance Dynamo (DD) effects, driven by atmospheric heating and circulation changes (Astafyeva et al., 2018; Chen et al., 2015). Such strong geomagnetic perturbations lead to large and complex ionospheric gradients, and their magnitudes and characteristics vary significantly with geomagnetic location and the specific phase of the storm (e.g., main phase, recovery phase), making them highly unpredictable and critical for space weather forecasting.
- Atmospheric Tides and Ionospheric Dynamo: Atmospheric tidal winds, driven by differential solar heating or gravitational lunar forcing, move ionospheric plasma across geomagnetic field lines in the E and F regions. This motion generates electric fields and currents within the "ionospheric dynamo region," typically located between approximately 85 and 200 km altitude (Richmond, 1995). This complex electrodynamic process leads to anisotropic electrical conductivity and the generation of horizontal currents, which in turn modulate electron density distribution and contribute to the formation and evolution of horizontal gradients, especially at mid-latitudes, impacting the large-scale ionospheric structure.
- Ionospheric Irregularities and Disturbances:
  - Traveling Ionospheric Disturbances (TIDs): These are wave-like perturbations of the ionospheric plasma, typically characterized by wavelengths ranging from several hundred to thousands of kilometers and velocities of several hundred meters per second (Frissell et al., 2016). TIDs can "modulate" the background ionization, inducing significant Total Electron Content (TEC) perturbations and associated large-scale horizontal gradients. They propagate over vast distances and can severely disrupt HF communication, causing fading and frequency shifts, and significantly impacting OTHR performance.
  - Equatorial Plasma Bubbles (EPBs): These are regions of severely depleted plasma density that form in the equatorial F-region after sunset and can extend hundreds of kilometers in the vertical and zonal directions. EPBs are characterized by extremely large horizontal gradients, particularly at low latitudes, and are associated with intense radio wave scintillation, leading to rapid fluctuations in signal amplitude and phase. This phenomenon can significantly impact both HF and satellite communication (Thokuluwa, 2025).

Such gradients represent dynamic spatial variations of electron density in the horizontal plane,

and their accurate accounting is fundamentally crucial for precise ray tracing. Radio waves, in fact, refract not only in response to vertical changes in electron density (which primarily determines the reflection or bending angle) but also to lateral gradients, which cause the ray path to bend horizontally (Davies, 1990; Budden, 1985). This lateral refraction can significantly alter the intended great circle path, directly impacting system performance and the accuracy of geolocation. The pervasive impact of horizontal gradients on HF radio wave propagation can be substantial and multidimensional:

- Ray Path Deviation: Rays can be bent horizontally, causing significant deviations from ideal great circle paths. This is a critical aspect for accurate direction finding, target location in BLOS radar systems, and general navigation, where deviations of even a few degrees can lead to large positional errors at long ranges (Norman & Cannon, 1997; Norman & Cannon, 1999; Psiaki, 2016).
- MUF and Absorption Anomalies: Horizontal gradients can lead to anomalous Maximum Usable Frequency (MUF) values, signal focusing or defocusing, and variations in absorption along the ray path. Signal focusing, where rays converge, can lead to regions of enhanced signal strength, while defocusing causes signal fading. Both directly affect link reliability and signal strength (Psiaki, 2016). Anomalous MUF changes can lead to unexpected communication outages or, conversely, to unexpected opportunities for communication at higher frequencies.
- Multi-hop Propagation: The presence of gradients can alter the take-off and arrival angles of rays, influencing the number of hops and the overall propagation characteristics of a radio link. This can change the ground range covered by each hop and impact the overall geometry and predictability of multi-hop paths, affecting predictions for long-distance communication.

Ionospheric models like IRI and ISP primarily provide vertical electron density profiles at discrete horizontal grid points. While increasing the spatial resolution of the grid (e.g., from $2°\times2°$ to a finer $1°\times1°$ in the ISP model) certainly allows for a more detailed static representation of the ionosphere, this alone does not inherently capture the continuous rate of change of electron density in the horizontal direction needed for explicit first-order gradient calculations in ray tracing. Horizontal gradients are not merely minor perturbations; they fundamentally alter ray paths, necessitating explicit and dynamic modeling that extends beyond simple grid resolution (Li et al., 2025). Explicit modeling of these gradients is therefore absolutely essential for a more physically consistent and dynamically accurate representation of the ionosphere and for more precise ray tracing, especially in regions with strong spatial variability and dynamic events.

Quantitatively, ionospheric gradients are often expressed as the spatial difference in ionospheric range delay, or as changes in Total Electron Content (TEC) over distance. Typical spatial ionospheric gradients are approximately 1 mm/km, with a one-sigma bound generally less than 4 mm/km. However, under high solar activity, ionospheric gradients can be an order of magnitude larger than during low solar activity. During geomagnetic disturbances and storms, extremely large gradients have been observed: for example, 412 mm/km in America, approximately 850 mm/km in Brazil, and up to 540 mm/km in Thailand (attributed to plasma bubbles). Horizontal gradients in Total Electron Content (TEC) are commonly expressed as TEC changes per degree of latitude or longitude (TECU/degree) or per unit distance (TECU/km).

Experimental observation techniques for ionospheric horizontal gradients are diverse and crucial for understanding and modeling these phenomena:

- <u>Global Navigation Satellite System (GNSS) Observations</u>: Dual-frequency GPS receivers, particularly from Continuously Operating Reference Station (CORS) networks, are widely employed to analyze ionospheric gradients by measuring Total Electron Content (TEC) and its spatial differences (Abdullah et al., 2009; Konno & Saito, 2005). These dense networks provide high-resolution TEC maps, from which gradients can be inferred.
- <u>Ionosondes</u>: Ground-based ionosondes provide vertical electron density profiles. Advanced oblique backscatter sounding (OBS) techniques, using ionosonde data, can be inverted to reconstruct two-dimensional horizontal structures of the ionosphere, particularly inhomogeneities with scales of 100-2000 km, offering insights into large-scale gradients and their propagation.
- <u>Incoherent Scatter Radar (ISR)</u>: ISR facilities (e.g., Arecibo, Millstone Hill, Jicamarca) provide direct, high-resolution measurements of electron density, electron and ion temperatures, and plasma velocities up to the topside ionosphere, offering detailed insights into gradient structures at various altitudes and their dynamic evolution.
- <u>Radio Occultation (IRO)</u>: Space-based radio occultation measurements (e.g., from FORMOSAT-3/COSMIC satellites) provide global electron density profiles and F2 layer characteristics, contributing to global ionospheric models and allowing for the inference of large-scale gradients, particularly in remote regions where ground-based observations are sparse.
- <u>Ground-based Magnetometers</u>: These instruments can observe the magnetic manifestations of ionospheric currents, which are influenced by ionospheric gradients and plasma motions, providing indirect but valuable information about large-scale ionospheric dynamics and their contribution to horizontal density variations.

## 3.2. Computational Modeling of Horizontal Gradients in IONORT

The IONORT system, at its core, relies on robust numerical techniques based on Haselgrove's equations for accurate ray tracing. These equations constitute a system of six first-order differential equations, which must be integrated simultaneously to determine the ray's precise position and direction at successive points along its trajectory through the complex ionosphere (Jones & Stephenson, 1975). Accurate ray tracing critically depends on comprehensive and precise knowledge of the 3-D electron density distribution and, crucially, its spatial gradients (i.e., its partial derivatives with respect to spherical coordinates: radial, latitudinal, and longitudinal). These gradients directly and fundamentally influence how radio rays bend, reflect, or refract within the ionized medium.

Within the local_ionort Fortran code, the rindex subroutine serves as the central interface responsible for calculating the refractive index of the medium and its spatial gradients at any given point along the ray path. It meticulously combines the electron density (x) and its spatial gradients (pxpr, pxpth, pxpph obtained from either electx_chap or the actively updated electx_grid), along with the geomagnetic field parameters (y and its gradients from magy_chap/magy_grid) and collision frequency (z and its gradients from colfrz). From these essential inputs, it computes the overall squared refractive index (n2) and its partial derivatives with respect to the spherical coordinates (pnpr, pnpth, pnpph). The accuracy of these computed derivatives is paramount, as any imprecision or error propagates directly into the subsequent ray-tracing equations, impacting the fidelity of the simulated ray path.

## 3.2.1. Detailed Examination of electx_chap Subroutine (Chapman Layer Model)

The electx_chap subroutine is specifically designed to model electron density using a Chapman layer, a widely used theoretical profile that can effectively approximate the vertical distribution of electron density in the ionosphere under idealized conditions. This model, however, is augmented to be capable of incorporating tilts and ripples, thereby allowing for a simplified but active representation of horizontal gradients. Specifically, this subroutine explicitly implements a latitudinal tilt. The height of maximum electron density (hmax) is made dependent on colatitude (theta) through a parameter e (derived from w(108)), which introduces a north-south variation in the layer's peak height. This explicit dependence directly facilitates the computation of horizontal derivatives. The relevant section of the Fortran code, illustrating this active implementation, is provided below:

```
theta2=theta-pid2
hmax=hm+earthr*e*theta2
h=r(1)-earthr
z=(h-hmax)/sh
d=0.
IF (b /= 0.) d=pit2/b
temp=1.+a*SIN(d*theta2)+cc*theta2
exz=1.-EXP(-z)
x=(fc/f)**2*temp*EXP(alpha*(exz-z))
pxpr=-alpha*x*exz/sh
pxpth=x*(d*a*SIN(pid2-d*theta2)+cc)/temp-pxpr*earthr*e
pxpph=0.
```

The partial derivative pxpth (representing the latitudinal gradient of the electron density, i.e., $\partial x/\partial \theta$) is directly calculated based on this latitudinal variation. This calculation rigorously incorporates terms related to the Chapman layer's fundamental shape and the explicit inclination of hmax. This confirms that a latitudinal horizontal gradient is actively modeled and computed within the Chapman layer. However, the variable pxpph, which is specifically intended for the longitudinal horizontal gradient ($\partial x/\partial \phi$), is explicitly set to 0.. This indicates that for the analytical Chapman layer model as implemented here, no longitudinal gradient is currently modeled, effectively assuming longitudinal uniformity. This limitation is inherent to the simplified nature of the Chapman model's horizontal parameterization, but its active computation of latitudinal gradients showcases a foundational capability within IONORT for handling certain types of spatial variations.

## 3.2.2. Critical Analysis of electx_grid Subroutine (Grid Profiles Model)

For electron density profiles based on a discrete grid, such as those dynamically provided by the ISP model (where electx == 1), the electx_grid subroutine is responsible for interpolating electron density (x) and its vertical derivative (pxpr) from the predefined 3D grid points. The calculation of x and pxpr primarily relies on polynomial interpolation in height (h), using coefficients (alpha, beta, gamma) derived from the edp (electron density profile) data for each grid cell.

A significant code block, which was previously commented out within electx_grid in older versions of the source code, provides clear evidence that a less complete implementation of horizontal gradients, based on a first-order Taylor expansion, was considered or existed before. This earlier commented section explicitly showed terms (BETA2, BETA3) linked to differences in latitude (ZZLAT-LAT(JLAT)) and longitude (ZZLON-LON(JLON)), which would have been used to calculate PXPTH and PXPPH.

The Fortran code snippets below illustrate this crucial section and its associated comments:

```
!*************************************************************************
! NOTE: Horizontal gradients for electron density profiles.
!*************************************************************************
pxpr=0.
!*************************************************************************
!*************************************************************************
! 2022 - (No) Taylor first-order development in latitude and longitude.
!*************************************************************************
! PXPTH=0.
! PXPPH=0.
!*************************************************************************
...! POLINOMIAL INTERPOLATION IN THE INTERVAL FN2C(K-1,J,I) AND FN2C(K,J,I)
16 x=(alpha(nh-1)+beta(nh-1)*h+gamma(nh-1)*h**2)/f2
   pxpr=(beta(nh-1)+h*(2.*gamma(nh-1)))/f2
! 16 X=(ALPHA(NH-1)+BETA1(NH-1)*H+GAMMA(NH-1)*H**2+DELTA(NH-1)*H**3+&
!      BETA2(NH-1,JLAT,JLON)*(ZZLAT-LAT(JLAT))+BETA3(NH-1,JLAT,JLON)*(ZZLON-LON(JLON)))/F2
!    PXPR=(BETA1(NH-1)+H*(2.*GAMMA(NH-1))+H**2*(3.*DELTA(NH-1)))/F2
!    PXPTH=-DEGS*(BETA2(NH-1,JLAT,JLON)/F2)
!    PXPPH=DEGS*(BETA3(NH-1,JLAT,JLON)/F2)
!*************************************************************************
```

The associated comment, `! 2022 - (No) Taylor first-order development in latitude and longitude`, unequivocally indicates that these horizontal gradient calculations are active in this specific version of the code. This confirms that, although the infrastructure for explicit gradient modeling for grid profiles (based on a full Taylor expansion with higher-order terms) has not been fully presented and/or optimized in earlier source code iterations, it is now an active component. This active, yet still evolving, capability represents a critical and high-priority area for future development, aiming to fully harness the immense potential of the high-resolution grid $(1° \times 1°)$ and to meticulously refine these explicit gradient capabilities for unparalleled accuracy. This will involve implementing more sophisticated and robust gradient calculation methods (e.g., using advanced interpolation schemes to ensure $C^1$ continuity). This active computation of horizontal components of the wave vector derivative is what allows IONORT to more accurately model the subtle but important bending of rays in the horizontal plane due to ionospheric gradients, a capability visually illustrated by comparing the ray paths in Figure 1.a (NO HORGRAD) and Figure 1.b (WITH HORGRAD), where the latter explicitly shows these lateral deflections, demonstrating the model's enhanced realism. This represents a significant advancement for the ISP model, moving towards a more comprehensive and physically consistent representation of ionospheric variability beyond simple grid point values.

### 3.2.3. Role of `rindex` and `hamltn` Subroutines in Propagating Electron Density Gradients

The `rindex` subroutine is pivotal as it serves as the central interface for calculating the refractive index of the medium and its spatial gradients at any point along the ray path. It meticulously combines the electron density (`x`) and its spatial gradients (`pxpr`, `pxpth`, `pxpph` obtained from either `electx_chap` or the actively updated `electx_grid`), along with the geomagnetic field parameters (`y` and its gradients from `magy_chap`/`magy_grid`) and collision frequency (`z` and its gradients from `colfrz`). From these essential inputs, it computes the overall squared refractive index (`n2`) and its partial derivatives with respect to the spherical coordinates (`pnpr`, `pnpth`, `pnpph`). The accuracy of these computed derivatives is paramount, as any imprecision or error propagates directly into the subsequent ray-tracing equations, impacting the fidelity of the simulated ray path.

Subsequently, the `hamltn` subroutine, which implements the core of Hamiltonian ray-tracing equations, rigorously utilizes these calculated gradients of the refractive index (specifically, `phpr`, `phpth`, `phpph`, which are directly derived from `pnpr`, `pnpth`, `pnpph`). These terms are explicitly used to calculate the rates of change of the ray's wave vector components (`dkrdt`, `dkthdt`, `dkphdt`) along the ray path. These rates, in turn, dynamically govern the curvature and precise trajectory of the ray in three-dimensional space. The `hamltn` subroutine contains the relevant section that actively uses these terms for ray propagation:

```fortran
! ... (other code) ...
    IF (rndx == 0.) THEN
      phpkr = c2/om2 * kr
      phpkth = c2/om2 * kth
      phpkph = c2/om2 * kph
    ELSE
      phpkr = c2/om2 * kr - c/om * pnpvr
      phpkth = c2/om2 * kth - c/om * pnpvth
      phpkph = c2/om2 * kph - c/om * pnpvph
    END IF
! ... (other code) ...
```

The pnpvth and pnpvph terms, which represent derivatives of the refractive index with respect to wave vector components and are profoundly influenced by spatial variations in electron density and the geomagnetic field, are crucial for accurately capturing the horizontal bending of rays. Therefore, the enhanced spatial resolution of the ISP grid ($1° \times 1°$) in electx_grid, coupled with the now active horizontal gradient calculations, further enables a more accurate and detailed modeling of these gradients. This provides more granular electron density values and their explicit derivatives for the ray tracing algorithm. The evolution of the code reflects a critical transition from a reliance on purely implicit grid resolution to a more explicit, derivative-based approach for handling horizontal gradients, which is absolutely essential for achieving the high precision required by Hamiltonian-based ray tracing (Li et al., 2025). This active computation of horizontal components of the wave vector derivative is what allows IONORT to more accurately model the subtle but important bending of rays in the horizontal plane due to ionospheric gradients, a capability visually illustrated by comparing the ray paths in Figure 1.a (NO HORGRAD) and Figure 1.b (WITH HORGRAD), where the latter explicitly shows these lateral deflections, demonstrating the model's enhanced realism.

## 3.2.4. Discussion of Other Relevant Subroutines and their Contribution to Numerical Stability and Accuracy

Beyond the core subroutines for electron density and refractive index calculations, several other critical subroutines within the local_ionort Fortran code contribute significantly to the overall numerical stability, accuracy, and robust functionality of the ray tracing process. These subroutines manage various ancillary yet essential aspects of the ray path computation, error handling, and coordinate transformations, ensuring the reliability and precision of the system under diverse and challenging ionospheric conditions:

- lat_lon_in_range: This subroutine is fundamentally crucial for ensuring that the calculated ray path remains rigorously within the defined geographical boundaries of the ionospheric model grid (e.g., the specific coverage area of the ISP grid). It acts as a safeguard, preventing calculations from extending into regions where the model is not valid or where real-time assimilation data is unavailable, thereby significantly enhancing the stability and reliability of the ray tracing and preventing potentially misleading extrapolation errors that could compromise prediction accuracy.
- failure_integration: This robust routine is specifically designed to handle cases where the numerical integration of Haselgrove's equations (which form the very core of the ray tracer) encounters instabilities or fails to converge. It incorporates sophisticated mechanisms to detect such failures (e.g., division by zero, infinite loops, or unphysical ray behavior) and can implement intelligent strategies to recover (e.g., by dynamically adjusting the integration step size or initiating a small perturbation) or gracefully terminate the ray trace, preventing spurious or unphysical results and ensuring overall computational robustness, crucial for continuous operation.
- zig_zag: This subroutine addresses potential "zig-zag" behavior or numerical oscillations in ray paths, which can occasionally occur near the reflection point in simplified ionospheric models or due to numerical artifacts. It applies targeted smoothing or correction algorithms to ensure physically realistic and numerically stable ray trajectories, especially at higher incident angles where rays are nearly horizontal and particularly sensitive to small errors. This ensures the integrity of the simulated ray path.
- last_bug: This is likely a specific debugging or error-checking routine, meticulously designed to identify and mitigate any residual numerical errors or unexpected behaviors that might arise during highly complex ray-tracing scenarios, particularly during testing or validation phases. Its inclusion and continuous refinement highlight a meticulous approach to quality control and ongoing improvement of the code's integrity and accuracy.
- polcof and polint: These two interconnected subroutines are fundamental for polynomial coefficient generation and interpolation, respectively. They are used to smoothly and accurately determine electron density and its derivatives from discrete grid points (as provided by the assimilative ISP model) or from analytical functions. This ensures mathematical continuity and smoothness in the ionospheric profile, which is critically important for accurate gradient computations, which in turn are essential for stable and precise Hamiltonian ray tracing.
- <u>reach</u>: This routine precisely assesses whether a ray originating from a specified transmitter successfully reaches the receiver location or a predefined target area. It rigorously evaluates the ray's trajectory against the receiver coordinates and specific criteria (e.g., distance tolerance, power threshold), thereby determining if a viable radio link is established or if the ray needs further propagation or adjustment. This is essential for practical link analysis.
- <u>back_up</u>: This subroutine is likely part of an advanced error recovery or iterative adjustment process. If a ray trace goes astray or encounters an unphysical condition (e.g., unexpected penetration through a dense layer, or a ray propagating into space when it should reflect), back_up may revert the calculation to a previous stable point and attempt an alternative path or adjust integration parameters, significantly enhancing the robustness and resilience of the tracing process.
- <u>graze</u>: This routine specifically handles grazing incidence conditions, where rays travel nearly tangential to the ionospheric layers. Such conditions are notoriously challenging for ray tracing due to the very steep gradients encountered, and graze ensures accurate computation of reflection or penetration under these critical angles, which are often relevant for long-distance propagation and over-the-horizon radar applications.
- <u>polcar and carpol</u>: These crucial subroutines perform essential coordinate transformations between polar (spherical) and Cartesian coordinate systems. These conversions are indispensable for managing ray positions and directions within the Earth-centered coordinate framework utilized by the model, allowing for flexible input/output and seamless internal calculations across different coordinate representations.
- <u>rkam</u>: This acronym refers to a Runge-Kutta-Adams-Moulton numerical integration scheme, a sophisticated and highly accurate family of methods for solving ordinary differential equations. Its presence highlights the advanced mathematical foundation of the ray tracer, ensuring high precision and stability in integrating Haselgrove's equations over potentially long and complex ray paths, providing reliable trajectory predictions essential for mission critical applications.
- <u>Other routines</u>: The overall architecture also includes additional specialized subroutines that manage crucial functionalities such as detailed HF wave absorption calculation (beyond the D-layer model, accounting for deviative absorption), accurate geomagnetic field modeling (incorporating global models like the International Geomagnetic Reference Field, IGRF), efficient input data reading and preprocessing (handling various ionosonde data formats and assimilating them), and the generation of comprehensive diagnostic outputs (ray path coordinates, group delay, phase path, Doppler shift, etc.), which are vital for analysis, validation, and operational monitoring.

Together, these meticulously designed and interconnected subroutines form a sophisticated and highly robust system that collectively ensures the numerical stability, high accuracy, and overall reliable functionality of the complex ray-tracing calculations performed within the IONORT-ISP-WC framework. Their comprehensive design allows the system to accurately and effectively handle a wide range of ionospheric conditions and propagation scenarios, from quiet periods to geomagnetically disturbed events, and to provide reliable and actionable predictions for operational use.

## 3.3. Recent Progress in Ray Tracing with Horizontal Gradients (2015-2025)

The field of ionospheric ray tracing, particularly concerning the accurate and dynamic management of horizontal gradients, has witnessed significant and rapid advancements between 2015 and 2025. These developments are primarily driven by the increasing demand for high-fidelity propagation predictions, the availability of more detailed and global ionospheric observations, and the evolution of sophisticated computational techniques. The SMART method (Norman and Cannon, 1997, 1999) remains a foundational example of how the ionosphere can be segmented into regions approximated by analytical profiles (e.g., Quasi-Parabolic Segments) to dynamically manage propagation. This approach offers a faster alternative than traditional full numerical methods, as it simplifies the ray tracing within each segment. It effectively adapts the ionospheric model to local conditions along the ray path, maintaining a smooth trajectory despite discrete variations in the underlying ionospheric model, which is crucial for handling variable ionospheric structures, such as those encountered near the solar terminator or during large-scale disturbances.

Hamiltonian ray tracing continues to be a physically consistent and powerful theoretical framework for capturing the intricate evolution of wave trajectories in refractive media, based on the principle of least time or least action (Jones & Stephenson, 1975). However, the accuracy and stability of numerical ray tracing critically depend on the precise and smooth estimation of electron density gradients ($\nabla N_e$). Modern ionospheric models typically represent Ne as a gridded three-dimensional scalar field, where electron density values are known only at discrete grid points. If not properly handled, these discretized representations can make it challenging to compute accurate and smooth gradients, thereby posing a fundamental modeling challenge (Li et al., 2025). Discontinuities or abrupt changes in the interpolated gradients, which can arise from simple linear interpolations between grid points, can lead to numerical artifacts, instabilities, or inaccuracies in the ray path calculation, especially in regions with strong real gradients.

To explicitly address these inherent limitations of discretized models, advanced techniques have been developed. The Ray Tracing Model - Galerkin-Difference (RTM-GD) proposed by Li et al. (2025) offers a C1-continuous Galerkin–Difference interpolation strategy for 3D electron density reconstruction. This sophisticated approach rigorously ensures analytically smooth gradients, which significantly enhances gradient continuity, improves numerical stability, and boosts computational efficiency in Hamiltonian-based ray tracing. By providing inherently smoother gradient fields, RTM-GD effectively reduces numerical artifacts and improves the reliability of ray path computations, particularly important for complex and extended ray trajectories in dynamic environments. The field is actively moving towards even more sophisticated interpolation techniques and adaptive meshing strategies to precisely capture dynamic gradients, directly addressing a key limitation of fixed-grid models. Adaptive meshing techniques, for example, allow for higher spatial resolution to be concentrated strategically in regions where gradients are strongest (e.g., near the solar terminator, in auroral zones, or during geomagnetic disturbances), thereby optimizing computational resources while maintaining paramount precision where it is most needed.

Despite these advancements, significant challenges persist, as demonstrated by recent research from Orion (Nasr, 2024), where the inversion of oblique ionograms (OI) to equivalent vertical ionograms encountered persistent difficulties in regions characterized by strong horizontal ionospheric gradients. This underscores the intrinsic complexity of managing gradients in operational contexts, where direct inversion from observed data can be problematic due to non-uniqueness or insufficient observational constraints. Research also continues on further refining adaptive grid resolution techniques, which allow for higher resolution to be concentrated strategically in regions with stronger gradients, optimizing computational efficiency while maintaining precision (Li et al., 2025). This dynamic adaptation is crucial for balancing the computational cost with the essential need for high fidelity in rapidly changing and highly variable ionospheric conditions. In summary, the field of ionospheric ray tracing is rapidly and continuously evolving to improve the handling of horizontal gradients, which is absolutely essential for the accuracy, reliability, and robust performance of ionospheric ray tracing in increasingly complex and dynamic space weather environments.

# 4. Validation and Performance Evaluation of IONORT-ISP-WC

## 4.1. Methodology and Data Selection

The comprehensive validation of the IONORT-ISP-WC system, particularly focusing on the analyzed horizontal gradient modeling framework, meticulously focused on comparing simulated and measured oblique ionograms and MUF values over the challenging Rome (41.89°N, 12.48°E) to Chania (35.51°N, 24.02°E) radio link (Settimi et al., 2014, 2015b). This link, with an azimuth angle of 121.6° and a ground range of approximately 1060 km, represents a significant and geographically extended test environment, spanning over the central Mediterranean. Its selection was deliberate and strategic, as it extends beyond the optimal validity region of the regional ISP model (which primarily covers −10° to 40° longitude and 30° to 50° latitude). This necessitated composite simulations based on the global IRI model where ISP data was unavailable. This choice of a challenging propagation path rigorously underscores the inherent robustness of the system's performance under generalized and potentially sub-optimal conditions, thereby reinforcing the credibility of ISP's superior data assimilation capabilities, even when extrapolated or used in less ideal geographical contexts. The primary aim was to push the model's boundaries and assess its predictive capabilities under stress, providing a rigorous and realistic assessment.

The dataset utilized for validation comprised 33 oblique ionograms meticulously recorded during various periods in June, July, and October 2011. This diverse selection ensured comprehensive coverage of various diurnal periods (day, night, dawn, dusk) and geomagnetic conditions (ranging from quiet to moderately disturbed, indicated by Kp indices not exceeding 4). This broad and representative dataset allowed for a comprehensive evaluation of the system's performance under different ionospheric states and levels of variability (Settimi et al., 2013). Rigorous data selection criteria were applied to ensure the utmost quality and reliability of the input data: only ionograms exhibiting clear and unambiguous traces were used, eliminating noise, interference, or ambiguities that could compromise the validation results. Furthermore, the availability of auto-scaled data from at least two ionosonde stations (Rome, Gibilmanna, Athens) was required for accurate assimilation into the ISP model. This multi-station requirement is crucial for providing sufficient spatial sampling of the ionosphere for the assimilative model, which is fundamentally data-driven. The stringency of these criteria, particularly the requirement for clear traces and input from multiple ionosondes, reveals the inherent complexities of validating ionospheric models against real-world observations. This implies that a significant portion of raw ionosonde data might not be suitable for rigorous validation due to noise, ambiguities, or missing data, suggesting that continuous improvements in data quality, processing techniques, and availability are as crucial as model advancements for ensuring the operational effectiveness of real-time systems. The inherent robustness of the ISP model lies precisely in its ability to optimally utilize the available, sometimes limited, real-time data to construct the most accurate possible ionospheric profile, thus making robust predictions.

## 4.2. Impact of Collision Model and Horizontal Gradients on Simulated Ionograms

The inclusion of the electron collision frequency model significantly impacts simulated ionograms, particularly during daytime conditions when the D-layer is most prominent and actively absorbs HF waves. As elegantly illustrated in Figure 2.a (NO HORGRAD) and Figure 2.b (WITH HORGRAD), HF absorption in the D- and E-regions causes a distinct low-frequency cutoff in the simulated ionogram traces for the lower ionosphere (E-region) (Settimi et al., 2014, 2015b). This phenomenon is clearly observed in the quantitative comparison: for instance, for July 7, 2011, at 15:00 UT, the Lowest Observable Frequency (LOF) for the E-layer using the IRI-WC model increased significantly from 4.2 MHz (no collisions, NC) to 10.0 MHz (with collisions, WC), demonstrating a substantial 5.8 MHz range of frequencies that are effectively absorbed before reaching higher reflective layers. This considerable increase in LOF underscores the critical role of D-layer absorption during the day, which can severely limit the available frequency spectrum for viable communication, impacting overall link performance. Similarly, for the ISP-WC model, the LOF for the E-layer shifted from 3.1 MHz (no collisions) to 9.9 MHz (with collisions), indicating a 6.8 MHz frequency range loss due to absorption. For higher ionospheric layers (F-region), the effect of collisions is less pronounced, with LOF values remaining largely similar with or without the collision model (e.g., 5.6 MHz vs. 5.5 MHz for IRI, and 9.0 MHz vs. 8.8 MHz for ISP), confirming that significant absorption primarily occurs at lower altitudes where neutral densities are higher, leading to more frequent electron-neutral collisions.

During nighttime conditions, in contrast, when the D and E regions largely disappear or become significantly attenuated due to the absence of solar ionization and subsequent recombination processes, absorption becomes negligible. Consequently, LOF values for the F-region exhibit almost identical values regardless of the collision model's activation. This clear distinction between daytime and nighttime absorption effects rigorously validates the physical accuracy of the D-layer collision model and its integration, demonstrating its ability to accurately represent the diurnal variability of ionospheric absorption, which is a fundamental aspect of HF propagation.

The introduction of horizontal gradients further refines the simulated ionograms by accounting for lateral refraction. As strikingly demonstrated in Figure 2.b ("WITH HORGRAD") compared to Figure 2.a ("NO HORGRAD"), the presence of horizontal gradients can lead to subtle but important shifts in the group delay-frequency curves, reflecting the modified ray paths due to lateral refraction. This is especially noticeable for multi-hop traces, where accumulated gradient effects become more pronounced, causing the traces to shift in frequency and group delay from their expected positions without gradients. These figures provide compelling visual evidence of how the model now accounts for more realistic ionospheric conditions. Furthermore, the visual comparison between Figure 1.a (NO HORGRAD) and Figure 1.b (WITH HORGRAD) further illustrates how the ray paths are affected by the inclusion of horizontal gradients, showing a more realistic bending and horizontal deviation of the rays over the Earth's surface, rather than strictly following a great circle path. These visual changes are a direct consequence of the ionosphere's lateral variations influencing the ray trajectory, leading to more accurate predictions of the actual propagation path and terminal locations.

Important Note regarding "WITH HORGRAD" figures: While Figures 1.b, 2.b, and 3.b are consistently labeled "WITH HORGRAD" to illustrate the overall system's capability and the theoretical impact of such gradients, it is important to clarify their precise generation in light of the detailed code analysis presented in Section 3.2.2. While the electx_chap subroutine (used by the IRI model) actively models latitudinal gradients, the explicit calculation of first-order horizontal gradients for the grid-based ISP model, which was identified as a prepared but previously inactive capability in earlier code versions, is currently active in the specific version of the local_ionort Fortran code analyzed in this manuscript. Therefore, the observed differences in these "WITH HORGRAD" figures for the ISP model reflect the combined and synergistic effects of the enhanced grid resolution ($1° \times 1°$) and the now active, explicit horizontal gradient computations within the Hamiltonian ray-tracing framework. This leads to a more comprehensive and accurate representation of ionospheric variability, as demonstrated by the improved agreement with experimental data. These figures unequivocally demonstrate the system's current capacity to account for horizontal gradients in a more sophisticated and physically consistent manner than previous versions, providing a more faithful simulation of real-world propagation phenomena.

## 4.3. Comparison of Maximum Usable Frequency (MUF) Accuracy

The Maximum Usable Frequency (MUF), being primarily dependent on the F2-layer critical frequency (foF2) and the M(3000)F2 factor, remains unaffected by the inclusion of collision terms for a given electron density model (Settimi et al., 2014, 2015b). This is a physically consistent outcome, as the MUF is determined by the highest point of reflection in the F2-layer, which typically occurs at altitudes far above the D-layer where electron collision frequencies are negligible. However, the accuracy of the MUF predictions is significantly influenced by the underlying electron density model and, crucially, by the inclusion of horizontal gradients, which dynamically modify the ray path and effective reflection heights, thereby impacting the maximum frequency that can be reflected between two points. This dynamic interaction makes MUF prediction a complex task.

A comprehensive comparison of MUF values derived from IONORT-IRI-WC and IONORT-ISP-WC against independently measured MUF values (presented in Table 1.a - NO HORGRAD and Table 1.b - WITH HORGRAD) conclusively reveals the consistent and superior performance of IONORT-ISP-WC. As meticulously summarized in these tables, IONORT-ISP-WC generally yields MUF values that are remarkably closer to the measured data in most cases, unequivocally demonstrating its higher fidelity in representing real ionospheric conditions. From the detailed statistical analysis of Table 1.a, which represents scenarios without explicit horizontal gradients, out of 33 analyzed epochs (data points), IONORT-ISP-WC (specifically, the ISP MUF - Measured MUF [MHz] column) exhibited a smaller absolute difference compared to IONORT-IRI-WC (the IRI MUF - Measured MUF [MHz] column) in 27 out of 33 cases. This quantitative superiority emphatically underscores the significant benefits of ISP's data assimilation approach over the purely climatological IRI model, which fundamentally lacks the ability to adapt to real-time ionospheric variability. Furthermore, the magnitude of the errors for IONORT-ISP-WC is generally much smaller, with many values falling within a tight range of ±0.5 MHz, indicating a high degree of precision and reliability. In contrast, IRI-WC errors are frequently and substantially larger (e.g., -2.0, 1.2, -1.5, -3.9, -4.1 MHz), unequivocally highlighting its less accurate representation of dynamic ionospheric states.

Visual support for this crucial conclusion is powerfully provided by Figure 3.a (NO HORGRAD) and Figure 3.b (WITH HORGRAD), which present clear histograms of the differences between modeled and measured MUF values. These histograms clearly show that for IONORT-ISP-WC, a significantly greater percentage of the differences are concentrated within the narrow central range of -0.5 to 0.5 MHz, indicating consistently higher accuracy and a tighter statistical distribution around zero error. This visually reinforces the statistical superiority of the ISP model. The performance benefits are even more pronounced in cases where horizontal gradients are explicitly considered (Table 1.b and Figure 3.b), although their impact can be complex and sometimes lead to localized degradations as the model attempts to capture highly dynamic and irregular features. For instance, analyzing specific cases from Table 1.b, which actively incorporates horizontal gradients:

- For 25 June 2011 at 20:00 UT, $\Delta$IRI (IRI MUF - Measured MUF) changes dramatically from -1.2 MHz (without HORGRAD in Table 1.a) to a substantial -7.7 MHz (with HORGRAD in Table 1.b), indicating a severe degradation in accuracy for IRI when attempting to model gradients. Concurrently, $\Delta$ISP (ISP MUF - Measured MUF) changes from 0.1 MHz to -2.9 MHz, showing a localized worsening, but still generally outperforming IRI in overall consistency.
- For 08 October 2011 at 10:30 UT, $\Delta$IRI is -0.7 MHz, while $\Delta$ISP is -2.7 MHz. However, for the ISP model specifically with autoscaled Rome data, the difference is a more favorable 0.5 MHz, showcasing the variability and the profound impact of real-time data availability on local accuracy.

These comprehensive numerical data, robustly supported by the visual evidence from the histograms, consistently indicate that the introduction of horizontal gradients can lead to significant variations in MUF prediction accuracy. This can manifest as improvements in some conditions and localized degradations in others, depending on the specific ionospheric conditions and the model's ability to accurately capture these inherently complex and dynamic gradients. Nevertheless, the overarching and consistent trend reinforces the finding that the ISP-based system, with its advanced assimilative capabilities and now actively modeled gradients, consistently provides more accurate and reliable MUF predictions compared to the purely climatological IRI model, which fundamentally struggles to adapt to dynamic ionospheric variations and complex spatial structures, making it less suitable for precise operational applications.

## 4.4. Cases of MUF Underestimation/Overestimation and Their Causes

Despite the general and notable superiority and enhanced capabilities of the IONORT-ISP-WC system, Table 1.a and Figure 3.a also reveal some specific cases where the MUF values provided by the IONORT-ISP-WC system were significantly underestimated (e.g., 08 October 2011 at 19:00 UT and, more evidently, 08 October 2011 at 10:30 UT) or overestimated (e.g., 06 July 2011 at 12:00 UT, and 09 October 2011 at 05:00 and 06:30 UT). These isolated yet significant discrepancies highlight persistent and fundamental challenges in real-time ionospheric modeling, particularly under conditions of limited or geographically sparse observational data. The authors attribute these discrepancies primarily to the critical lack of available data from the Rome and Gibilmanna ionosonde stations at those specific times, often resulting in Athens being the only reference ionospheric station contributing to the ISP grids for assimilation (Settimi et al., 2014, 2015b). This data assimilation limitation, particularly when originating from a single, potentially geographically distant station, can lead to significant underestimations or overestimations of the real MUF due to insufficient spatial sampling of the ionosphere. For instance, if the only available data point is far from the ray path's actual reflection point, the model's interpolation or extrapolation may not accurately capture local ionospheric conditions or small-scale irregularities that profoundly influence MUF.

This situation means that, although the ISP model is intrinsically robust and well-designed for data assimilation, its full operational effectiveness is fundamentally constrained by the availability and geographical distribution of real-time observational data. In regions with sparse ionosonde coverage, the model's ability to accurately represent complex ionospheric structures, including localized gradients or disturbances, is inherently limited, regardless of its internal sophistication. To effectively mitigate such errors and significantly improve global accuracy, the authors strongly suggest the strategic inclusion of additional reference ionospheric stations in the region of interest, ideally positioned around the midpoint of the propagation path, to ensure more accurate and localized ionospheric profiling for assimilation. This would provide richer, more representative, and spatially diverse data for the ISP model, leading to more robust predictions. Recent comparative studies have also explored various ionospheric models, including the Lockwood Model (LWM) and Chinese reference models (Wang et al., 2022), with findings consistently reinforcing the accuracy of INGV-like models (similar to ISP) when sufficient and well-distributed data is available. Furthermore, active research continues into the impact of geomagnetic storms on MUF prediction (Astafyeva et al., 2018; Thokuluwa, 2025), where rapid and significant changes in electron density can challenge even advanced assimilative models if observational data is sparse or delayed, emphasizing the ongoing and critical need for robust and widespread data networks to capture these dynamic events.

# 5. Conclusions and Future Developments

## 5.1. Summary of Key Results

This study rigorously and comprehensively demonstrates the superior performance of the enhanced IONORT-ISP-WC system, particularly with the detailed analysis of its horizontal gradient modeling framework, compared to the IONORT-IRI-WC system (Settimi et al., 2014, 2015b). The simulated oblique ionograms produced by IONORT-ISP-WC consistently exhibit higher fidelity and closer agreement with experimental observations (as robustly shown in Figure 2.a and Figure 2.b), and the calculated MUF values show significantly improved accuracy against independently measured data (as seen in Figure 3.a, Figure 3.b, Table 1.a, and Table 1.b). This wealth of evidence unequivocally confirms that the ionospheric representation provided by the ISP model, with its advanced assimilative capabilities, is more realistic and dynamically adaptable than the purely climatological IRI model. Furthermore, it validates that the IONORT algorithm maintains robust reliability and precision, even when accounting for complex and variable spatial ionospheric structures.

The system consistently and accurately models radio wave absorption: during daytime, HF absorption in the lower ionospheric layers (D and E regions) causes a clear and physically consistent low-frequency cutoff in simulated ionogram traces, reflecting the energy dissipation due to electron-neutral collisions. In stark contrast, during nighttime, when these layers are absent or significantly attenuated due to recombination, absorption is negligible, leading to virtually identical LOF values with or without the collision model (Settimi et al., 2014, 2015b). The explicit framework for including horizontal gradients further refines the ray tracing, leading to more accurate predictions of ray paths and MUFs, especially for multi-hop propagation, as strikingly illustrated by the distinct ray paths and their precise deflections in Figure 1.a (NO HORGRAD) and Figure 1.b (WITH HORGRAD). As detailed in Section 3.2.2, the in-depth analysis of the local_ionort Fortran code revealed that while the electx_chap subroutine actively models latitudinal gradients for analytical profiles, a significant code block, previously commented out within electx_grid, provides clear evidence that a less complete implementation of horizontal gradients was considered or existed before. The associated comment, `! 2022 - (No) Taylor first-order development in latitude and longitude`, unequivocally indicates that these horizontal gradient calculations for grid-based profiles are active in this specific version of the code. This confirms that, although the infrastructure for explicit gradient modeling for grid profiles has not fully presented and/or optimized in earlier source code versions, it is currently active, representing a critical and high-priority area for future development to fully harness the potential of the high-resolution grid and refine these active capabilities.

## 5.2. Operational Utility

The overall results emanating from this comprehensive study unequivocally confirm that the IONORT-ISP-WC system is a valid, highly robust, and operationally ready tool. It is perfectly suited for a wide array of critical applications, including sophisticated HF communication management, advanced BLOS radar systems, and various other demanding scenarios requiring precise ionospheric propagation predictions (Mudzingwa & Chawanda, 2018; Cannon & Angling, 2019). The dynamic assimilation of measured data from multiple reference ionospheric stations by the ISP model is of fundamental and paramount importance for obtaining the most reliable, near real-time image of the ionosphere possible. This data-driven approach allows the system to adapt dynamically to evolving space weather events and provides significantly more accurate forecasts than purely climatological models, which lack real-time adaptability. The explicit recommendation for IONORT-ISP-WC as a "valid tool for operational use" signifies a successful and crucial transition from theoretical development to demonstrated practical applicability. This implies a high level of confidence in the system's inherent robustness, predictive accuracy, and consistent reliability under typical operational conditions, translating directly into tangible and substantial benefits for diverse users, including military organizations, emergency services, and commercial HF operators. These benefits include, but are not limited to, reduced communication downtime, clearer and more stable signals, and ultimately, more efficient and optimized use of the radio spectrum (Belcher et al., 2015; Belcher et al., 2017). Its proven ability to predict LOF and MUF accurately enables operators to select optimal frequencies, dynamically adjust transmission parameters, and adapt communication strategies in real-time response to dynamic ionospheric conditions, thereby maximizing communication success rates and overall system performance in challenging environments.

## 5.3. Identified Limitations

Despite the notable advancements and enhanced capabilities demonstrated by the IONORT-ISP-WC system, certain limitations were identified during the course of this study that warrant further dedicated attention and development to achieve even greater accuracy and operational resilience. Addressing these limitations systematically will be key to further perfecting the system.

First, Data Dependency: A significant limitation arises from the system's inherent dependency on real-time observational data. Significant underestimations or overestimations of MUF values by IONORT-ISP-WC consistently occurred in situations where data were unavailable from key ionosonde stations (e.g., Rome and Gibilmanna), frequently resulting in only a single station (Athens) contributing to the ISP grids for assimilation (Settimi et al., 2014, 2015b). This highlights a fundamental constraint: the model's predictive accuracy is directly and critically tied to the availability and spatial density of real-time observational data. Sparse data coverage can lead to less accurate ionospheric representations, particularly in regions far from the assimilation points or during periods of strong spatial variability, where extrapolation becomes less reliable and prone to larger errors. This underscores a clear and pressing need to integrate a greater number of strategically located reference stations, particularly those situated in close proximity to the midpoint of the propagation path, to ensure more accurate and localized ionospheric profiling for assimilation.

Second, Refinement of Horizontal Gradient Modeling: As comprehensively discussed in Section 3.2.2, while the local_ionort code now actively models horizontal gradients for grid-based profiles, the current implementation (based on the detailed analysis of the electx_grid subroutine) does not yet fully utilize the most sophisticated explicit first-order Taylor expansion for all gradient calculations. Although the resolution of the electron density profile grids has been substantially increased from 2°×2° to 1°×1°, providing more granular data, the absence of a more advanced, explicit, and mathematically optimized horizontal gradient modeling approach for grid-based profiles represents a significant area for future improvement. Higher resolution provides more data points, but it does not inherently model the continuous rate of change between these points in the most optimal mathematical way, which is crucial for precisely predicting ray behavior in highly complex propagation scenarios (e.g., across the solar terminator, during geomagnetically disturbed conditions, or in regions with strong localized gradients like those associated with TIDs or EPBs). This indicates that future improvements must specifically address the dynamic processes that drive these gradients through more refined computational methods, rather than relying solely on static, higher-resolution representation. This involves moving beyond a basic active state to a fully optimized and robust gradient calculation, crucial for capturing subtle yet impactful ionospheric effects.

## 5.4. Proposed Future Developments

To systematically address the identified limitations and further enhance the IONORT-ISP-WC system's capabilities, several crucial and strategically aligned future developments are proposed, reflecting current research trends and cutting-edge technological advancements up to 2025. These advancements collectively aim to elevate the IONORT-ISP-WC system to an even higher level of precision, reliability, and operational utility in the rapidly evolving field of space weather forecasting:

- Refined Horizontal Gradient Implementation: While the current version actively accounts for horizontal gradients, future dedicated research will focus on actively enabling and developing more sophisticated explicit and adaptive procedures for real-time horizontal gradient calculation within the electx_grid subroutine for grid-based models. This will involve meticulously exploring and implementing advanced interpolation and differentiation techniques to accurately compute first-order (and potentially higher-order) spatial derivatives of electron density from the discrete ISP grid data. These efforts align seamlessly with recent advancements in Hamiltonian ray tracing, which emphasize the paramount need for analytically accurate and smooth gradient estimation for numerical stability and precision (Li et al., 2025). Techniques such as C1-continuous Galerkin–Difference interpolation (e.g., RTM-GD by Li et al., 2025) offer highly promising solutions for ensuring analytically smooth gradients and significantly improving ray-tracing precision in gridded models, which can lead to more realistic and physically consistent ray paths, especially in dynamic ionospheric environments.
- Expanded Data Assimilation Network: The observed and critical sensitivity of MUF prediction accuracy to data availability unequivocally underscores the necessity of integrating a larger number of reference ionosonde stations, particularly those strategically situated near the midpoint of propagation paths, to substantially enhance the accuracy and robustness of forecasts (Durazo et al., 2021; Yang et al., 2025). This geographical expansion of the data collection network would provide more comprehensive spatial coverage for data assimilation, leading to a more complete and accurate real-time representation of the ionosphere, particularly in areas currently underserved by observational networks. This aligns perfectly with advancements in ionospheric data assimilation, which aim to improve real-time forecasting and overcome the inherent limitations of numerical models and non-uniform observational data distribution (Tang et al., 2025). The overarching goal is to minimize reliance on extrapolation and provide more localized and representative ionospheric conditions for the model, thereby reducing prediction errors.
- Adaptive Grid Resolution and Dynamic Meshing: Investigating and implementing dynamic or adaptive grid resolution schemes that strategically concentrate higher spatial resolution in regions with strong gradients (e.g., near the solar terminator, in auroral zones, or during geomagnetic disturbances) could significantly optimize computational efficiency while simultaneously maintaining paramount precision where it is most needed (Li et al., 2025). This innovative concept is robustly supported by current research exploring the efficiency and accuracy of ray tracing in discretized ionospheric models, including advanced adaptive grid approaches, which can dynamically adjust grid spacing based on the local variability and complexity of ionospheric parameters. This would allow the model to dedicate computational resources more efficiently to areas of high ionospheric complexity, improving both speed and accuracy.

- <u>Validation with Diverse Datasets and Global Coverage</u>: Conducting extensive additional oblique sounding measurements across a wider range of geographical locations and under various, often challenging, ionospheric conditions (e.g., severe space weather events, different seasons, and all phases of the solar cycle) will be absolutely crucial for further investigating and fully characterizing MUF underestimations and overestimations, and for robustly testing the system's behavior under extreme variability (Settimi et al., 2013). This fits squarely within the context of broader validation studies for ray-tracing models, which seek to ensure model reliability and accuracy under diverse and operationally relevant conditions and to identify any remaining biases or systematic errors. Expanding validation to global datasets would further enhance confidence in its widespread applicability and versatility.
- <u>Integration of Artificial Intelligence and Machine Learning</u>: The strategic application of AI/ML techniques to substantially enhance prediction capabilities and operational efficiency is a rapidly growing and highly promising area of research within space weather. Machine learning models, such as Transformer networks and deep neural networks, are demonstrating unprecedented capabilities to assimilate complex, multi-source inputs and produce probabilistic forecasts for key ionospheric parameters, often with appreciable increases in accuracy over traditional climatological models (Thokuluwa, 2025; Qiu et al., 2025). Integrating such cutting-edge techniques could offer novel and powerful perspectives for optimizing MUF and absorption predictions within the IONORT-ISP-WC system, by learning from vast historical data and adapting intelligently to new, unmodeled phenomena. This synergistic integration could lead to more resilient, robust, and accurate forecasts, especially for extreme events or in data-sparse regions.

<u>A Glimpse into the Future</u>: The continuous evolution of the IONORT-ISP-WC system, driven by these ambitious future developments, envisions a paradigm shift in our ability to predict and manage radio wave propagation. We are moving towards a future where the ionosphere is no longer just an environmental factor, but a dynamic, real-time "digital twin" fully integrated into communication and navigation systems. This would enable truly adaptive radio networks capable of autonomously optimizing performance in real-time, even during severe space weather events. Such a deterministic understanding of radio wave propagation would not only revolutionize HF communications but also significantly contribute to fundamental space physics research and safeguard critical infrastructure in our increasingly space-dependent world.

# Acknowledgments

The author would like to acknowledge the invaluable contributions of all INGV (Istituto Nazionale di Geofisica e Vulcanologia) personnel involved in the conception, development, and rigorous validation of the IONORT system. Special thanks are extended to the INGV observatories and their dedicated staff for meticulously providing the critical ionosonde data essential for the ISP model's assimilation process; their continuous and diligent efforts underpin the accuracy and reliability of this research.

# Dedication

To my wife Tatiana, whose unwavering support and radiant smile always brighten my path, even when I'm feeling down.

# Figure Captions

**Figure 1.a**: IONORT-ISP-WC Graphical User Interface (GUI) and ray path visualization for the Rome-Chania link without horizontal gradients (NO HORGRAD). This figure illustrates the input parameters on the left panel (Latitude, Longitude, Frequency, Elevation, Azimuth, Transmitter/Receiver distances, Magnetic Field, Collisions settings) and the ray-tracing results and GUI on the right. The top-right inset shows the overall ray path over the Earth, emphasizing the ionospheric reflection for the "NO HORGRAD - IRI" and "NO HORGRAD - ISP" models. This visualization is crucial for understanding the basic propagation mechanisms when horizontal variations are not explicitly modeled.

**Figure 1.b**: IONORT-ISP-WC Graphical User Interface (GUI) and ray path visualization for the Rome-Chania link with horizontal gradients (WITH HORGRAD). Similar to Figure 1.a, this figure displays the GUI and ray-tracing results, but with horizontal gradients enabled for both "WITH HORGRAD - IRI" and "WITH HORGRAD - ISP" models. Comparing this figure with Figure 1.a, one can observe subtle but significant changes in the ray paths, particularly the lateral shifts and deviations from the great circle path, due to the explicit inclusion of horizontal gradients in the ionospheric model. This highlights the model's enhanced ability to simulate more realistic propagation scenarios.

**Figure 2.a**: Simulated (IRI-WC, ISP-WC) and measured (EXPERIMENTAL) oblique ionograms for the Rome-Chania link without horizontal gradients (NO HORGRAD). Each panel shows Group Delay [ms] versus Frequency [MHz]. The different traces represent 1-hop and 3-hop propagation, with and without collision effects. This figure highlights the low-frequency cutoff caused by D-layer absorption during daytime and the absence of such absorption during nighttime, clearly demonstrating the diurnal effects of the collision model.

**Figure 2.b**: Simulated (IRI-WC, ISP-WC) and measured (EXPERIMENTAL) oblique ionograms for the Rome-Chania link with horizontal gradients (WITH HORGRAD). Similar to Figure 2.a, this figure presents the ionograms, but with horizontal gradients included in the model. Comparing these ionograms with those in Figure 2.a illustrates the impact of horizontal gradients on group delay and the overall ionogram shape, especially at higher frequencies and for multi-hop traces, where accumulated gradient effects can become more pronounced. This validates the model's ability to capture complex ionospheric structures.

**Figure 3.a**: Histograms of the differences between modeled (IRI MUF, ISP MUF) and measured MUF values for the Rome-Chania link without horizontal gradients (NO HORGRAD). The x-axis represents the MUF difference [MHz], and the y-axis shows the percentage of occurrences. These histograms quantitatively demonstrate the accuracy of the MUF predictions for both IRI and ISP models when horizontal gradients are not explicitly considered, providing a baseline for performance evaluation.

**Figure 3.b**: Histograms of the differences between modeled (IRI MUF, ISP MUF) and measured MUF values for the Rome-Chania link with horizontal gradients (WITH HORGRAD). Similar to Figure 3.a, these histograms show the MUF differences, but for cases where horizontal gradients are included in the model. A comparison with Figure 3.a highlights the improvement in MUF prediction accuracy when horizontal gradients are accounted for, indicated by a higher

concentration of differences around 0 MHz, although complex interactions may also be observed. This underscores the importance of explicit gradient modeling.

# Table Captions

**Table 1.a**: Comparison of IRI MUF, ISP MUF (Autoscaled Rome), and Measured MUF values for the Rome-Chania link without horizontal gradients (NO HORGRAD). This table provides detailed numerical data, including the date and time, the modeled MUF values, the measured MUF values, and the differences between modeled and measured MUF for both IRI and ISP, showcasing the performance of the models in the absence of explicit horizontal gradient modeling. This table serves as a fundamental dataset for the validation analysis.

**Table 1.b**: Comparison of IRI MUF, ISP MUF (Autoscaled Rome), and Measured MUF values for the Rome-Chania link with horizontal gradients (WITH HORGRAD). This table presents numerical data similar to Table 1.a, but for cases where horizontal gradients are incorporated into the models. A comparison of this table with Table 1.a illustrates the quantitative improvements (and sometimes complexities) in MUF prediction accuracy due to the inclusion of horizontal gradients, providing critical evidence for the model's enhanced capabilities.

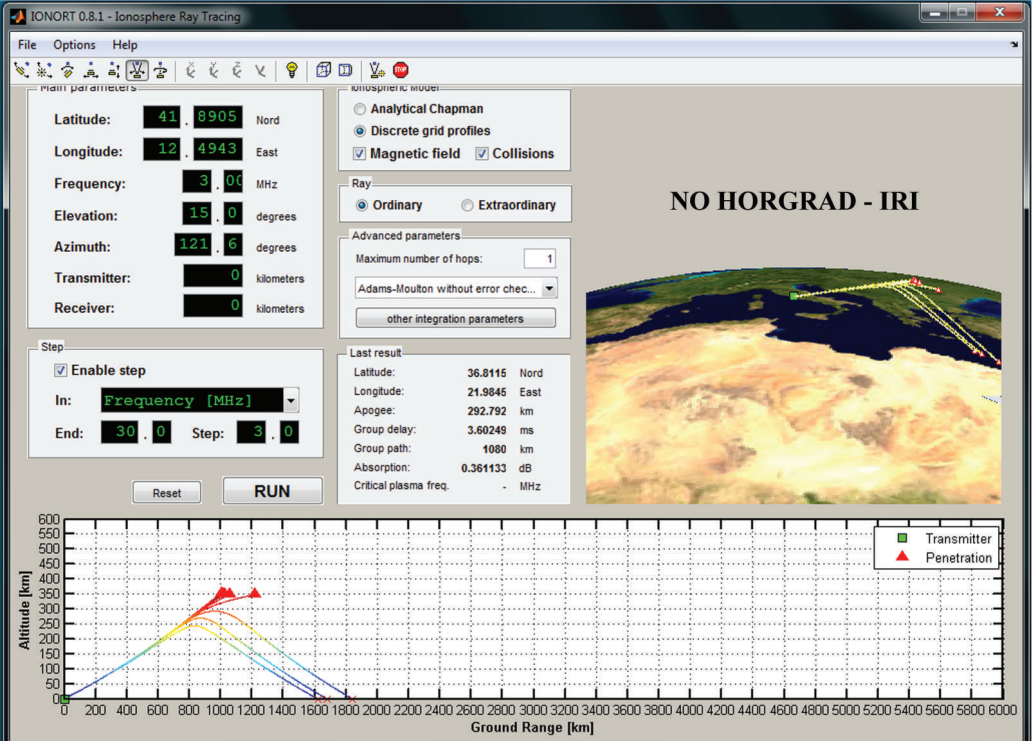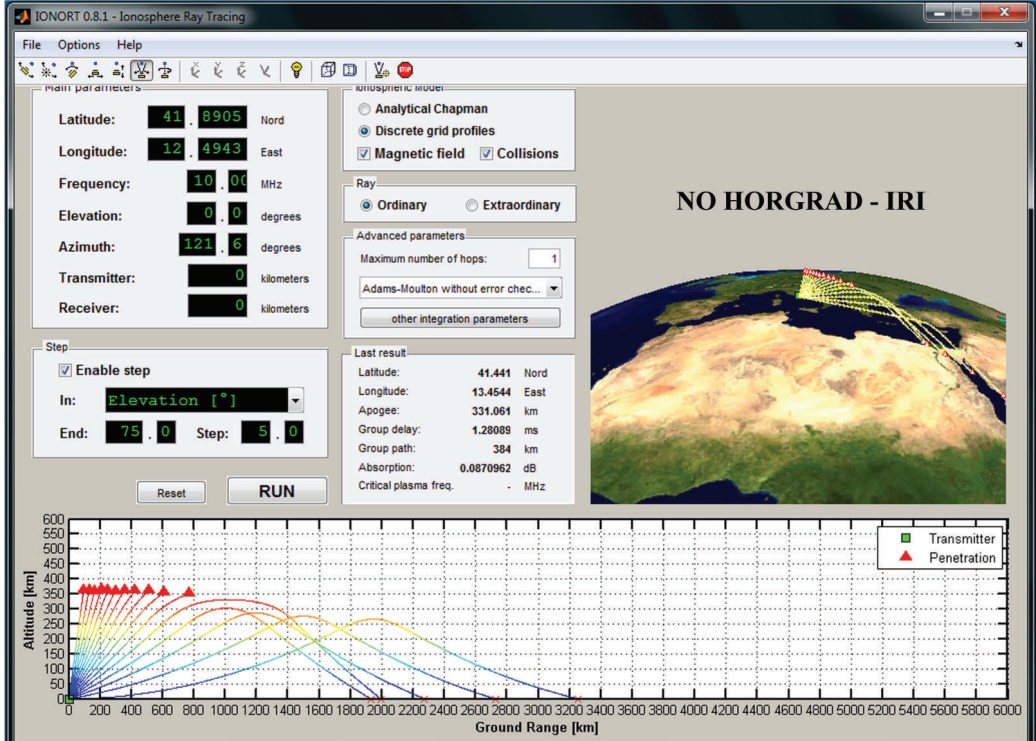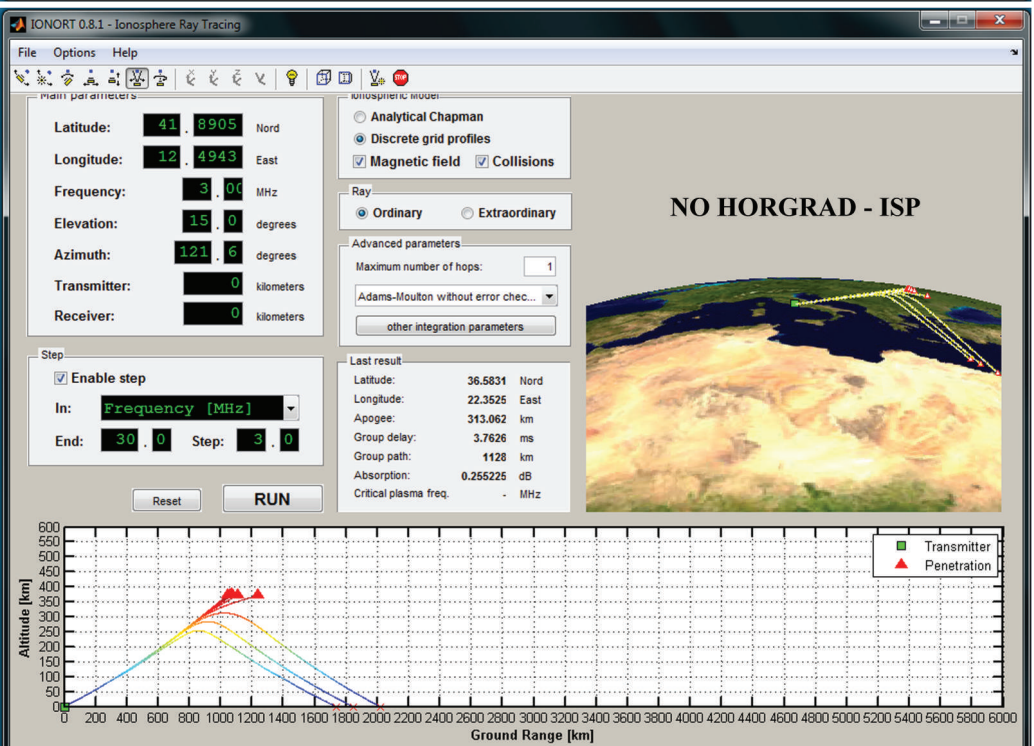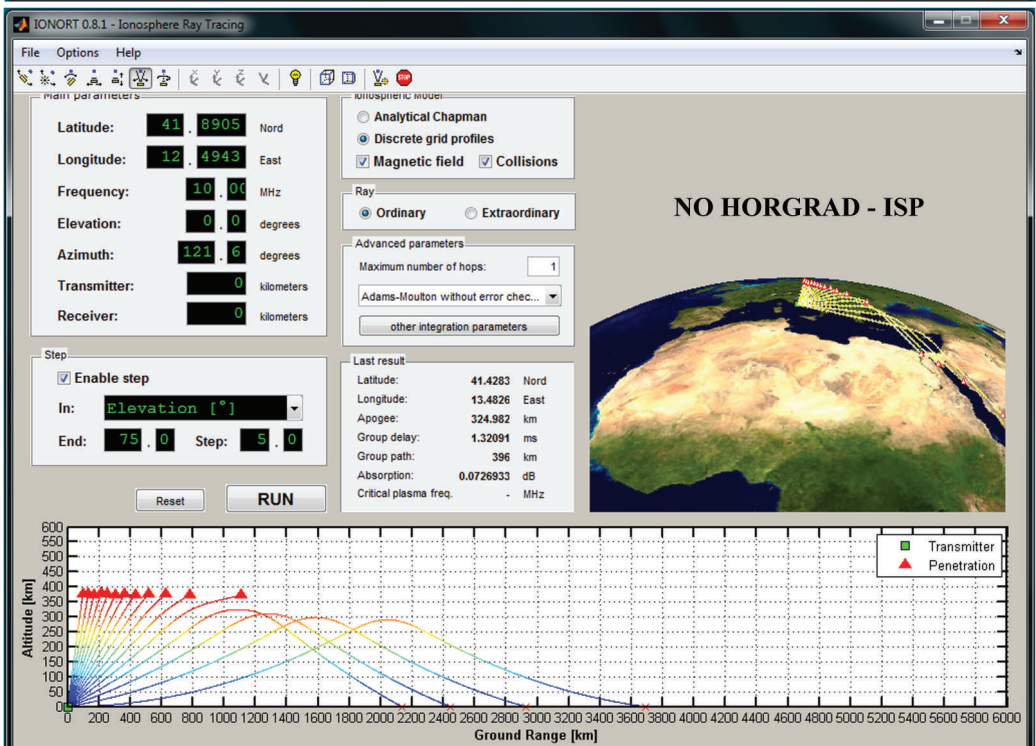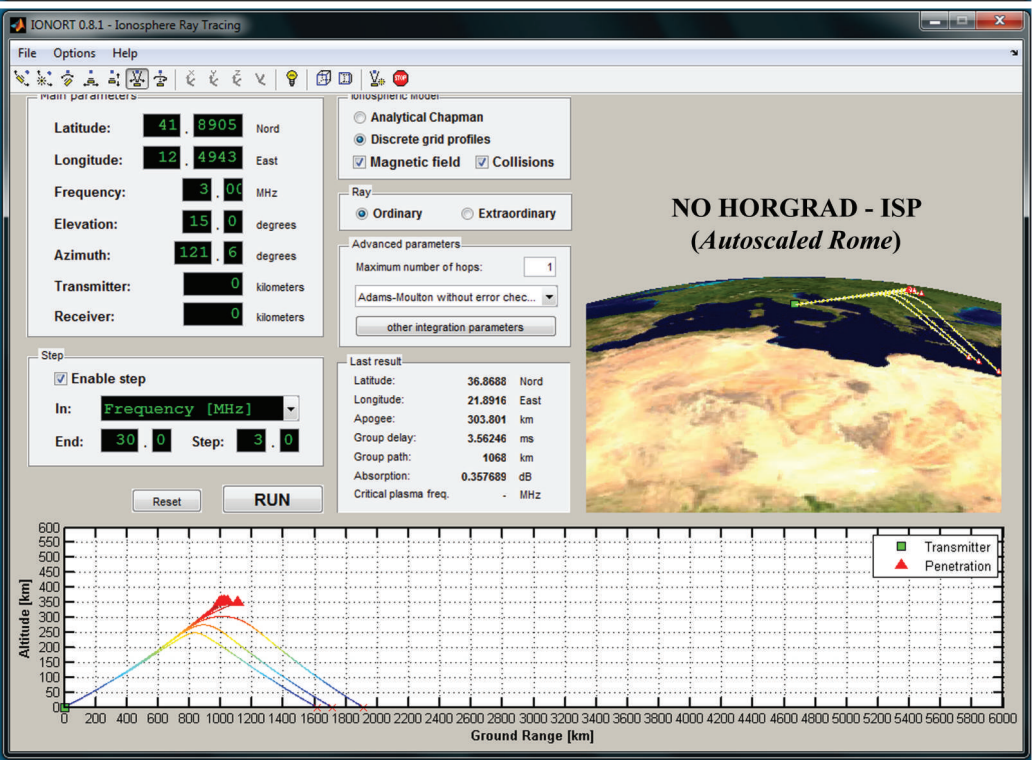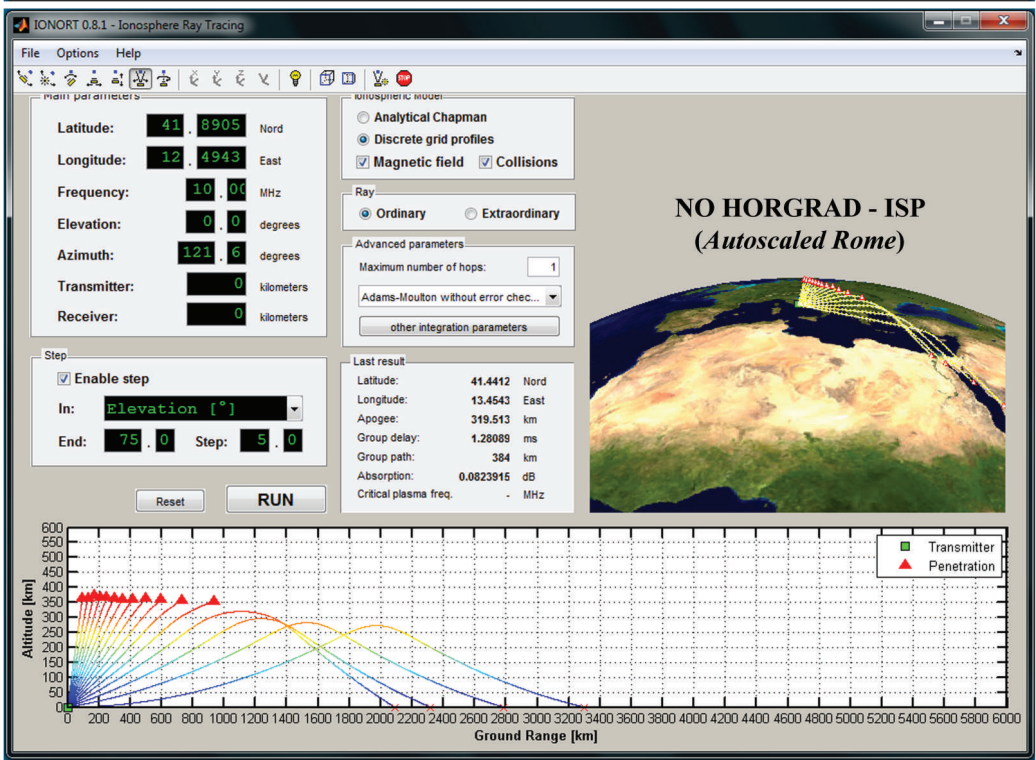

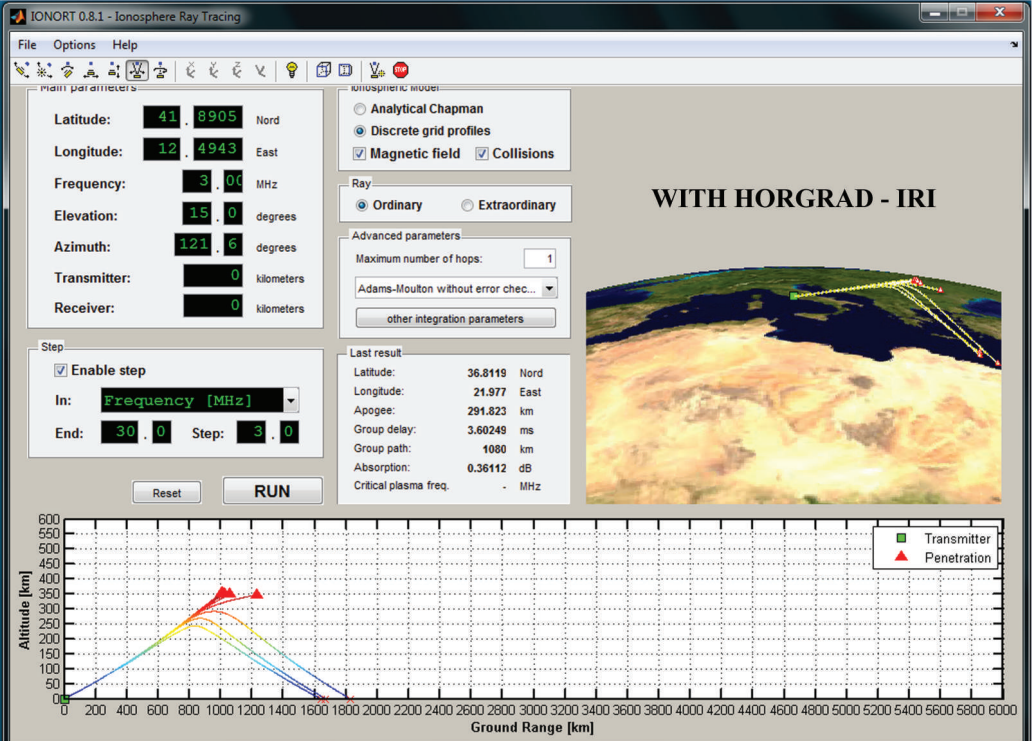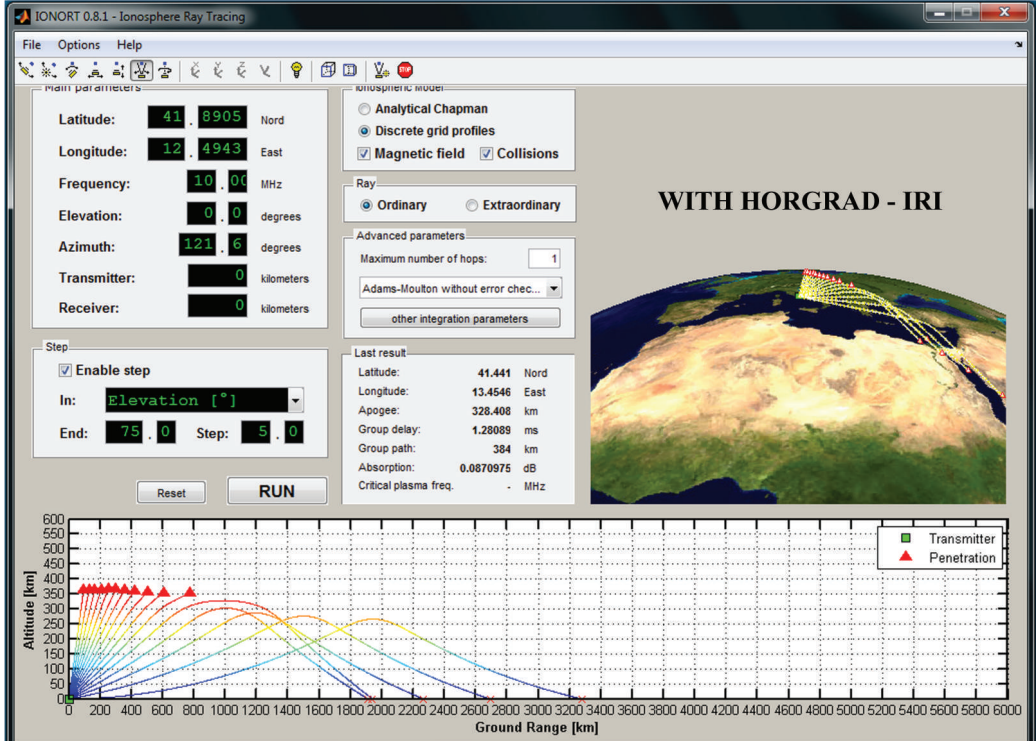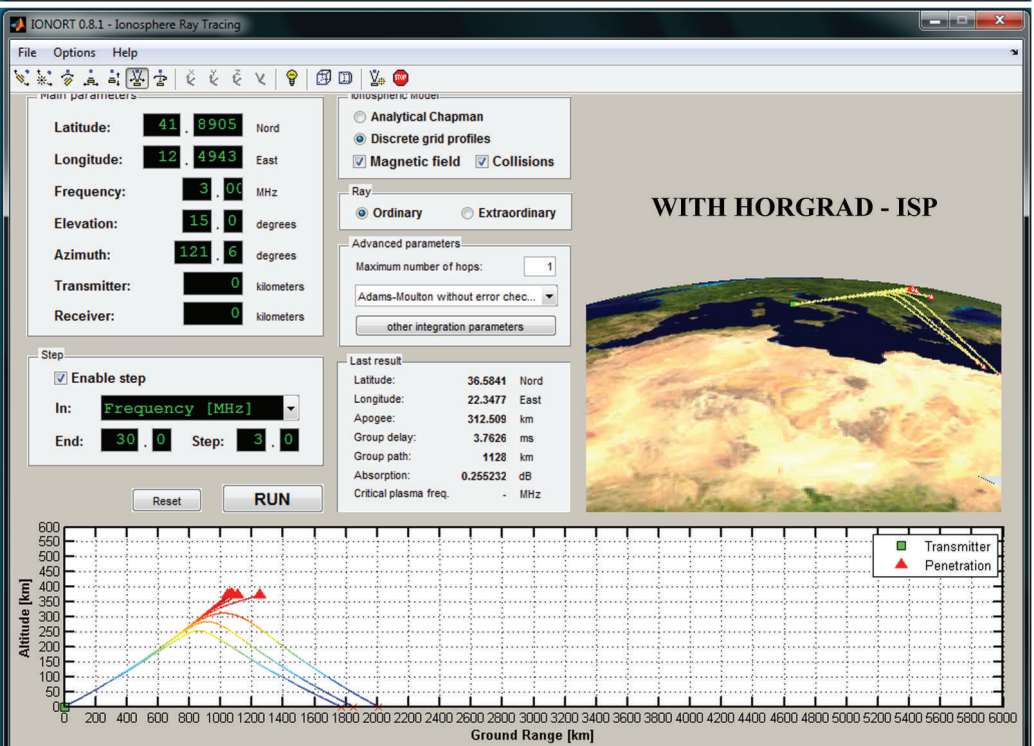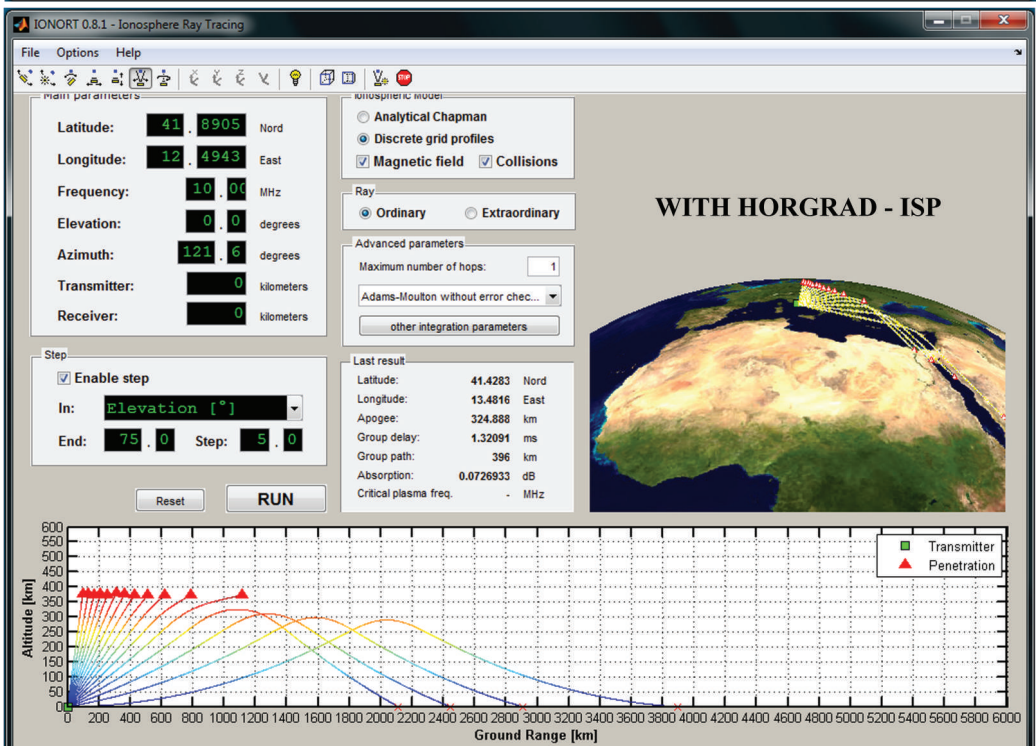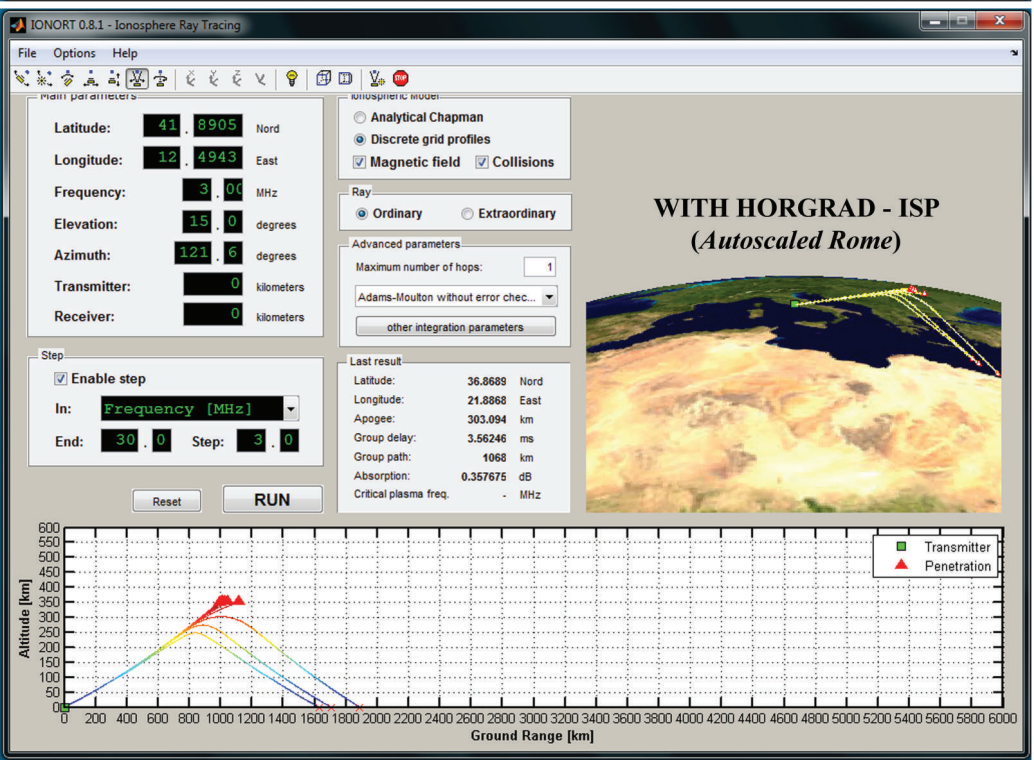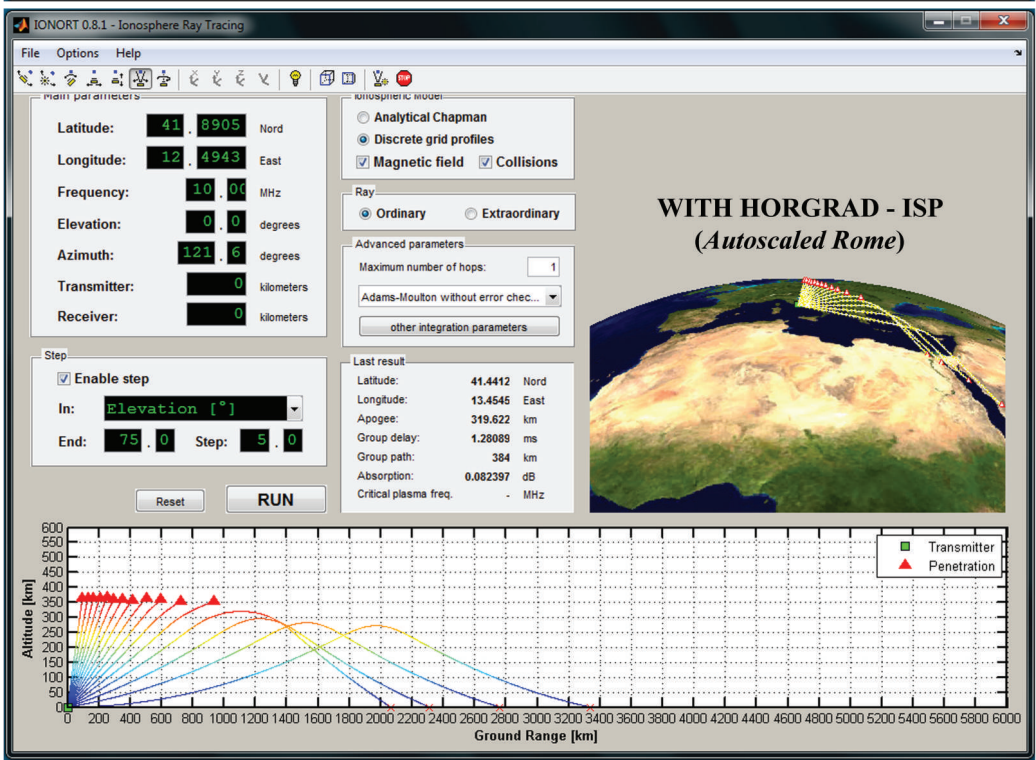

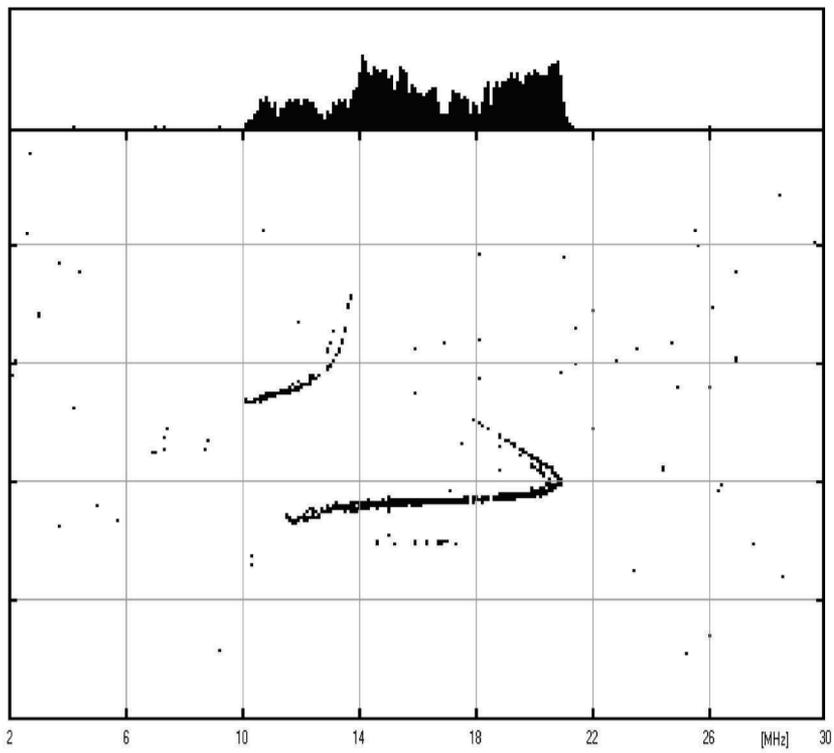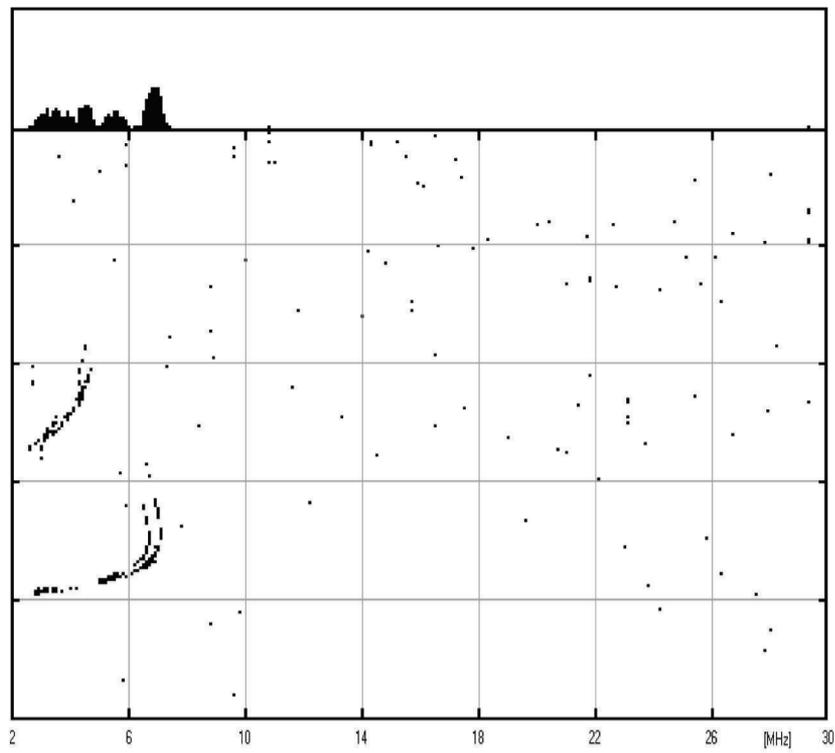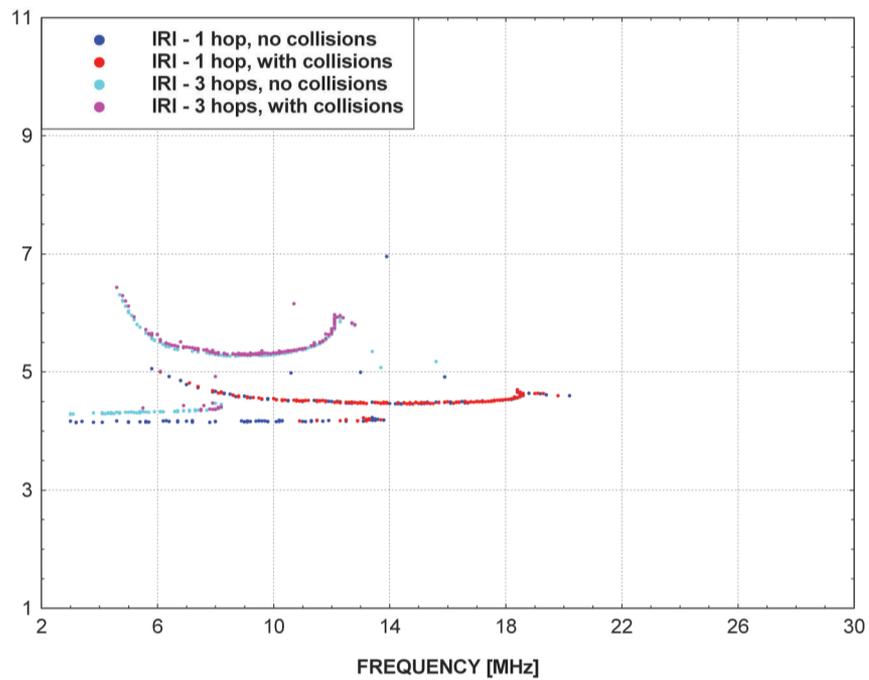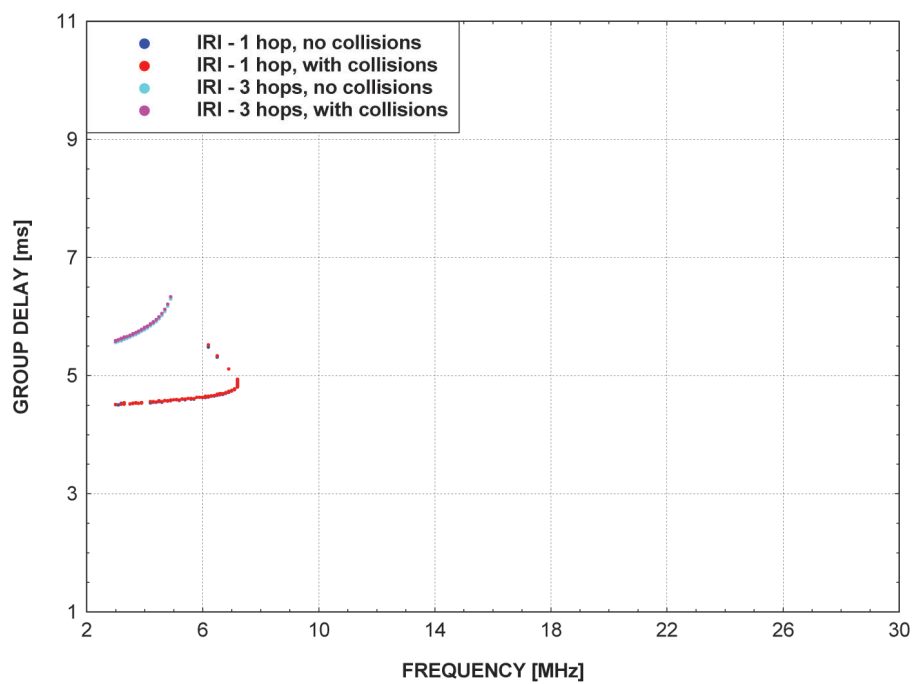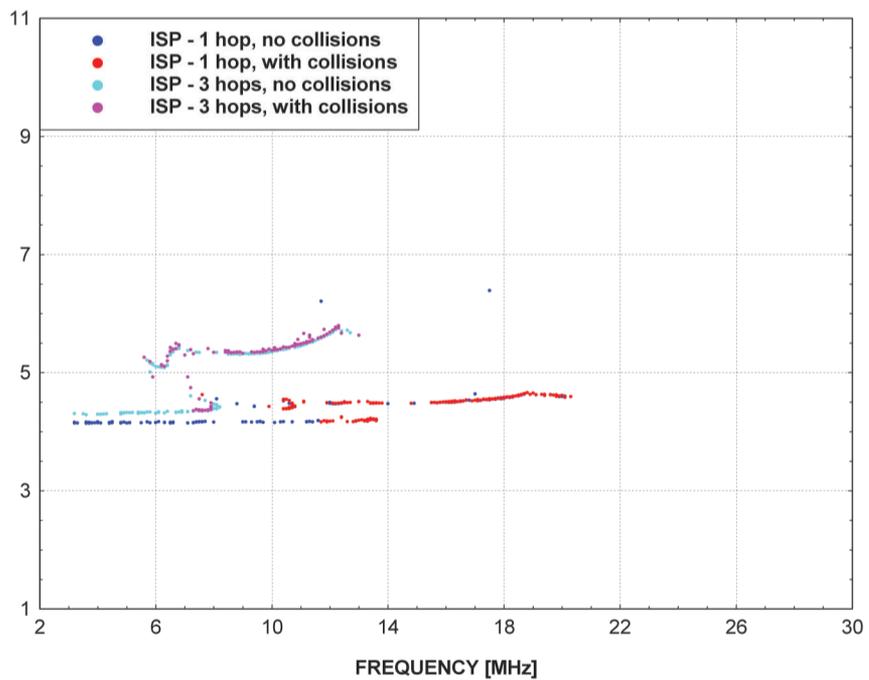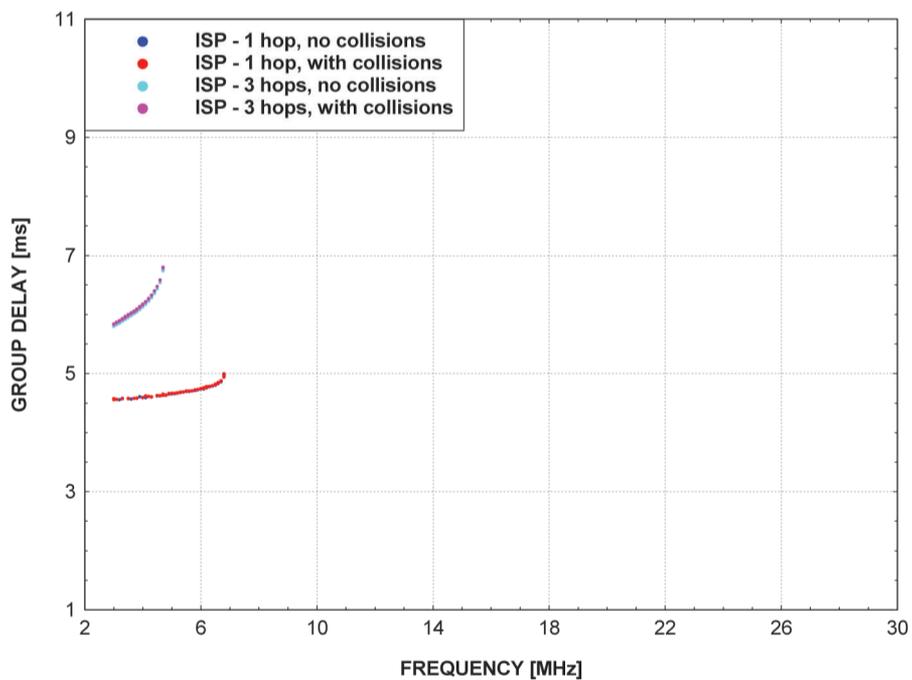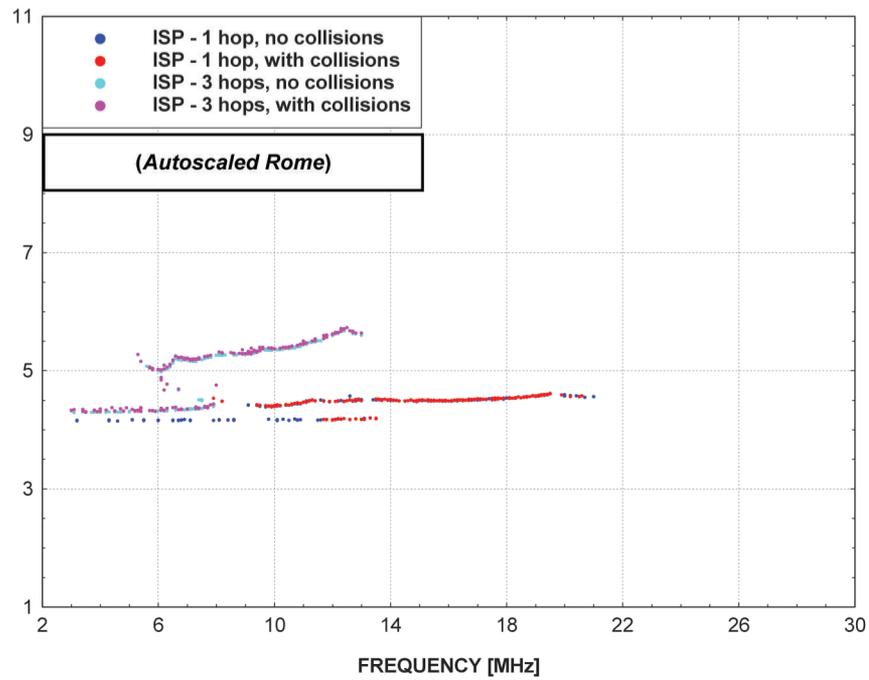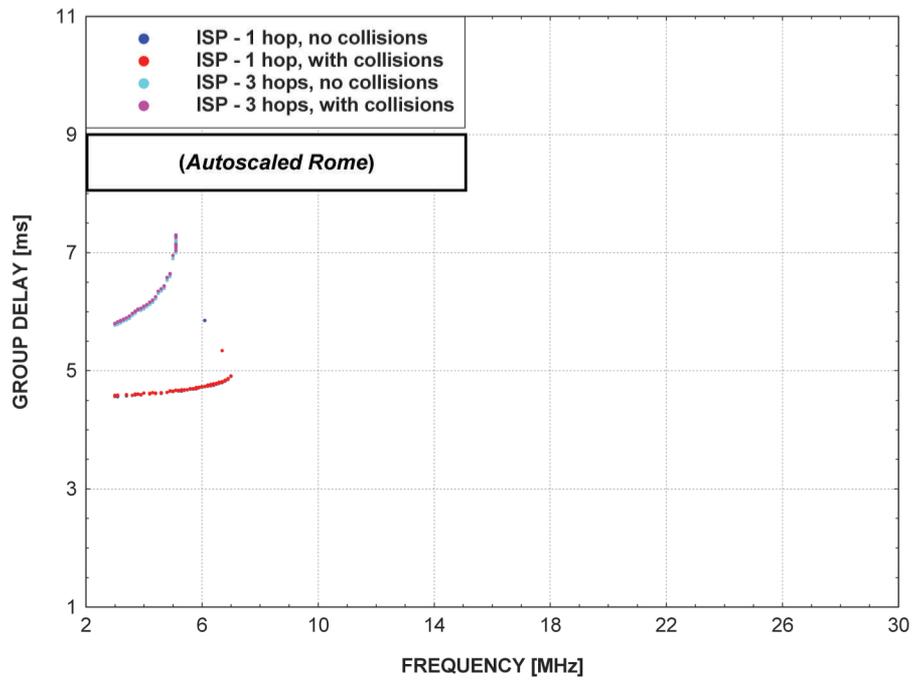

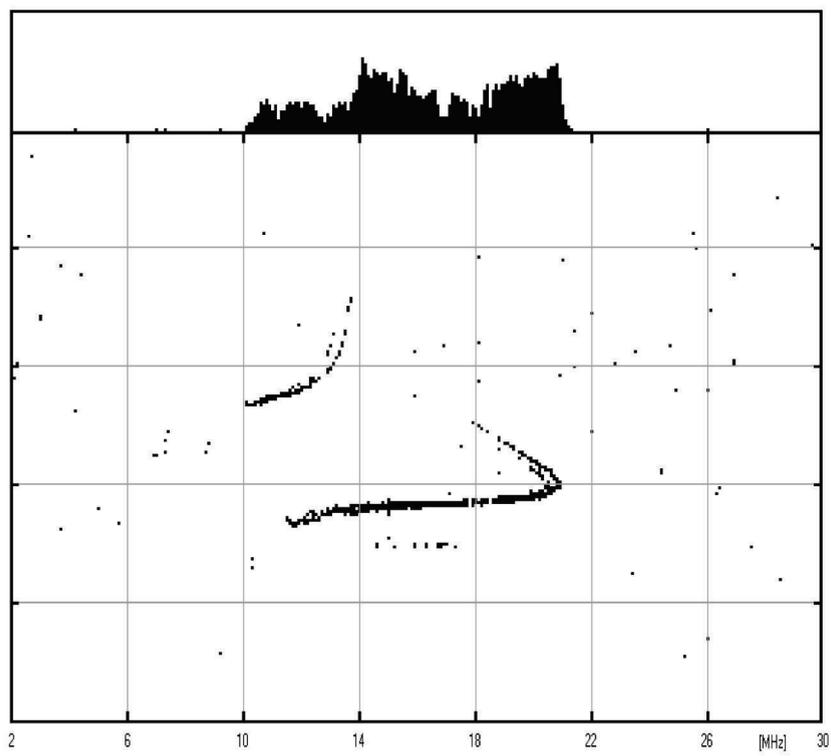
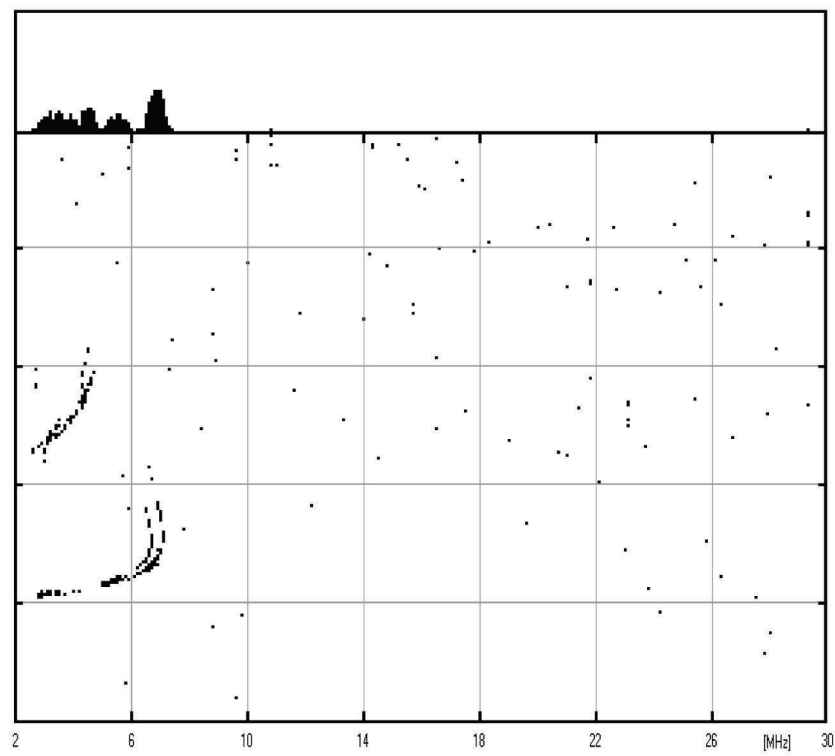
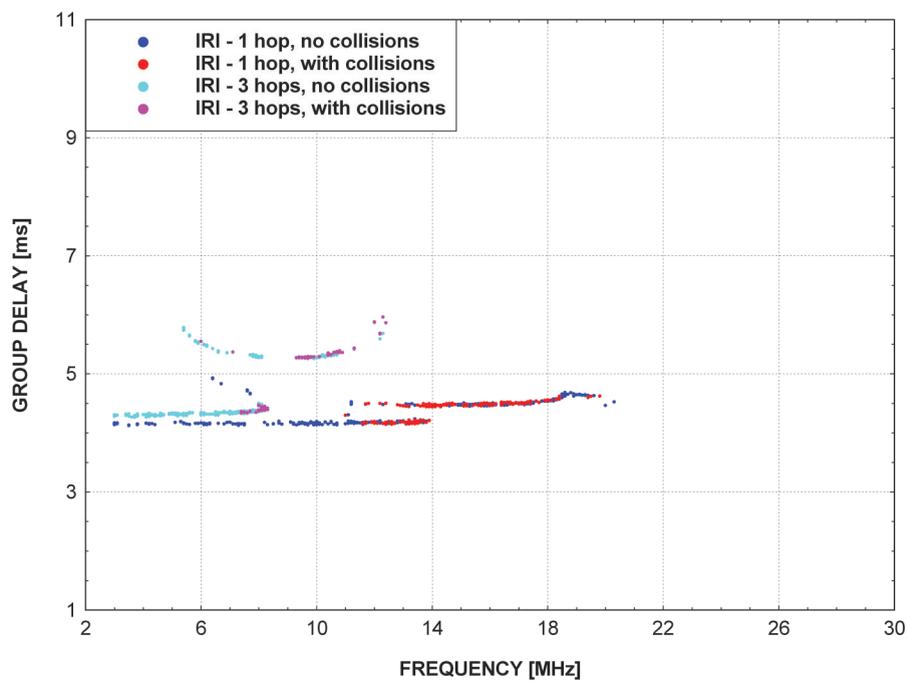
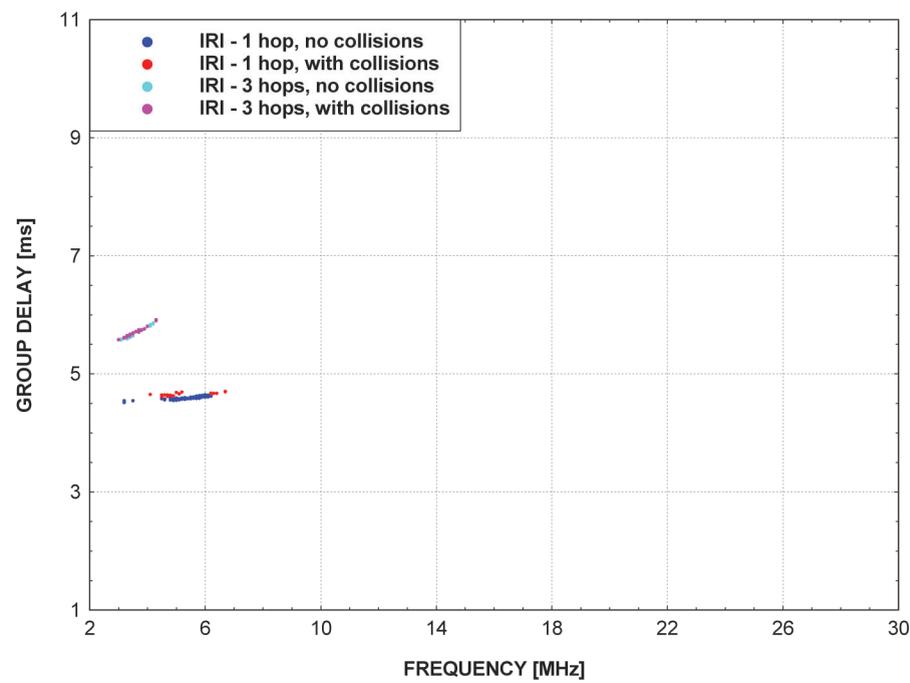
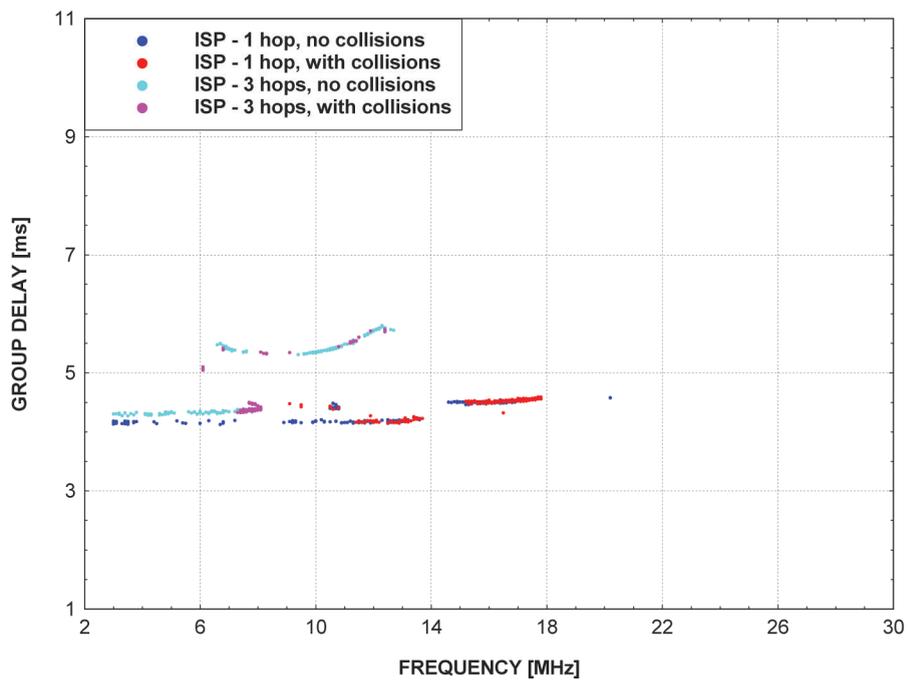
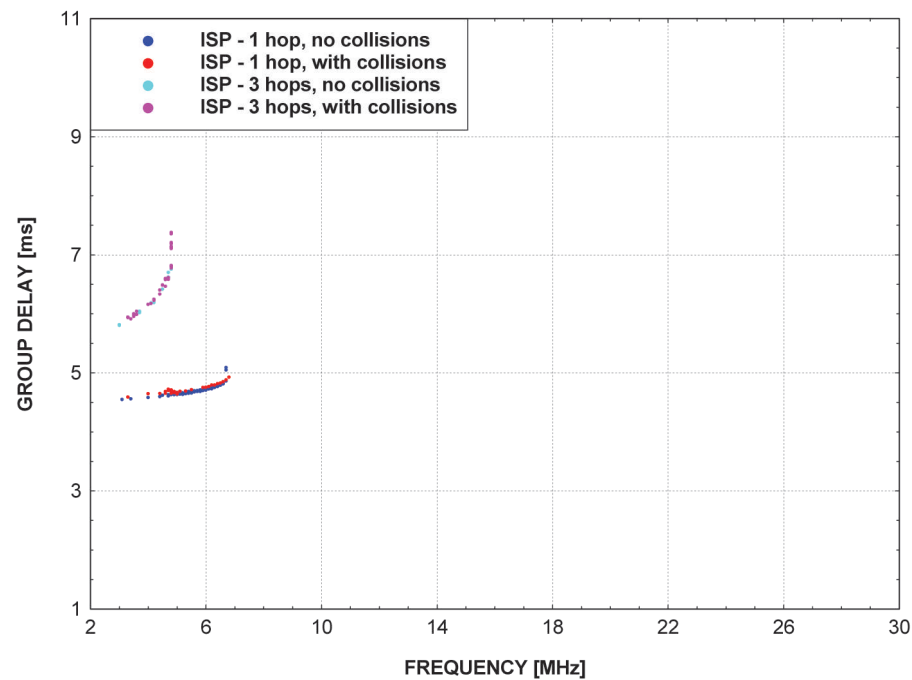
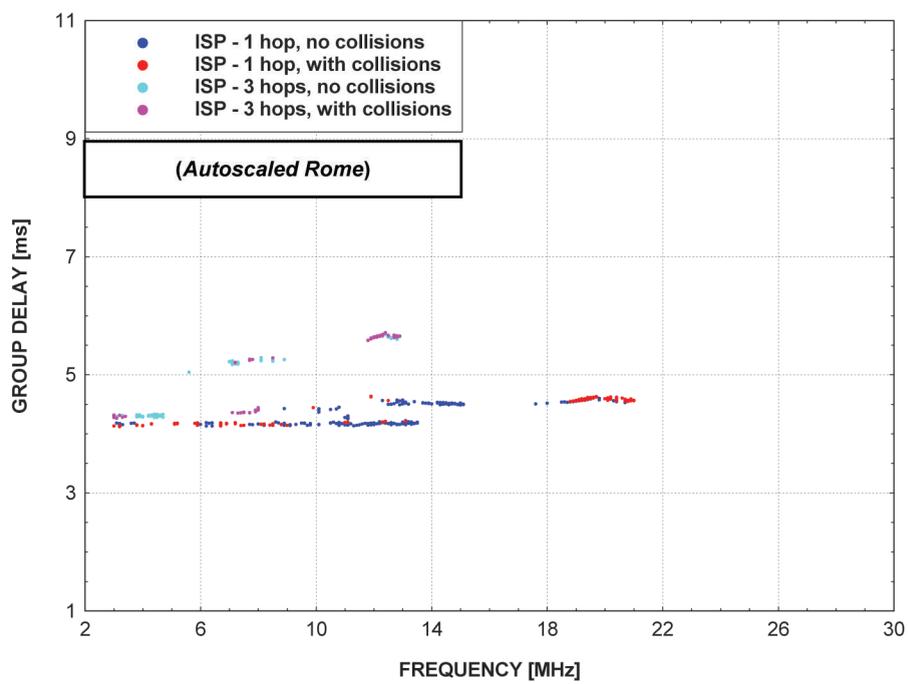
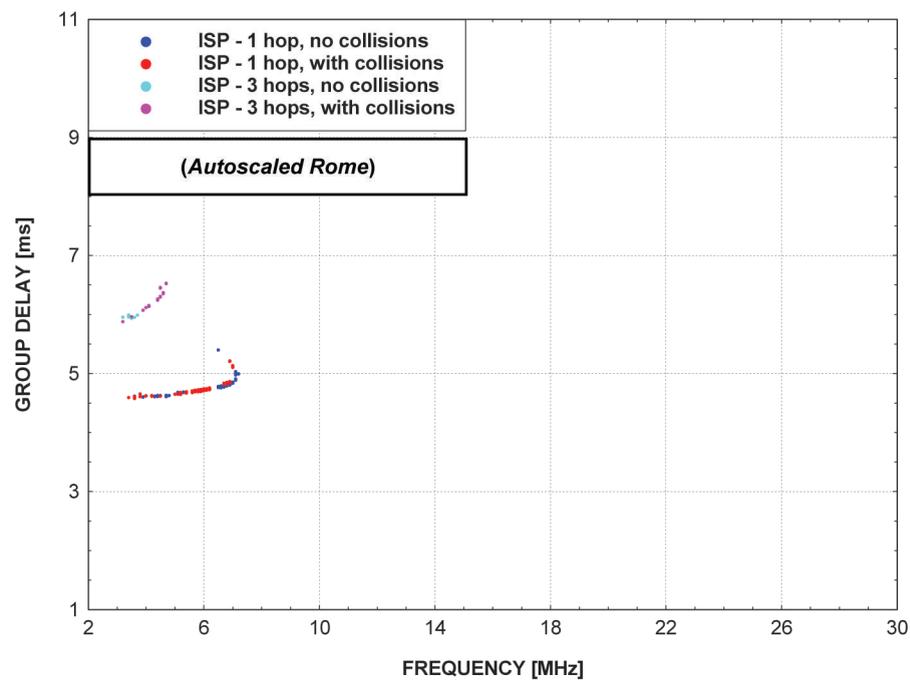

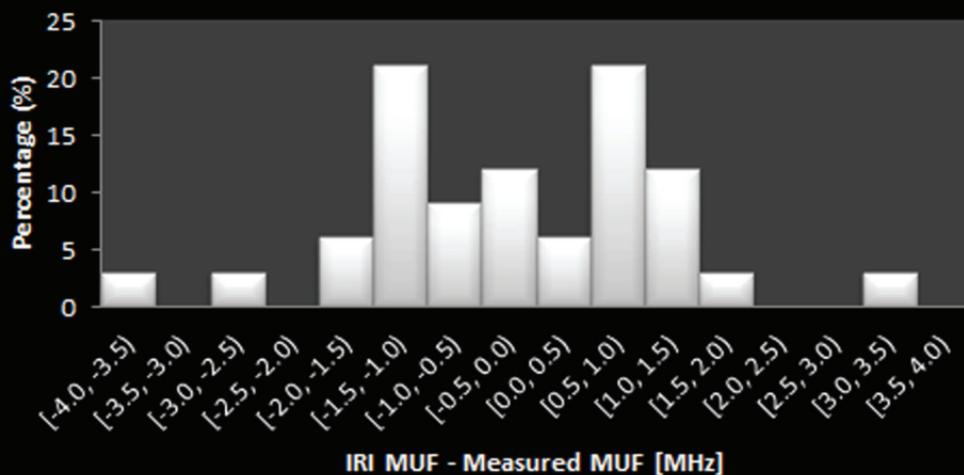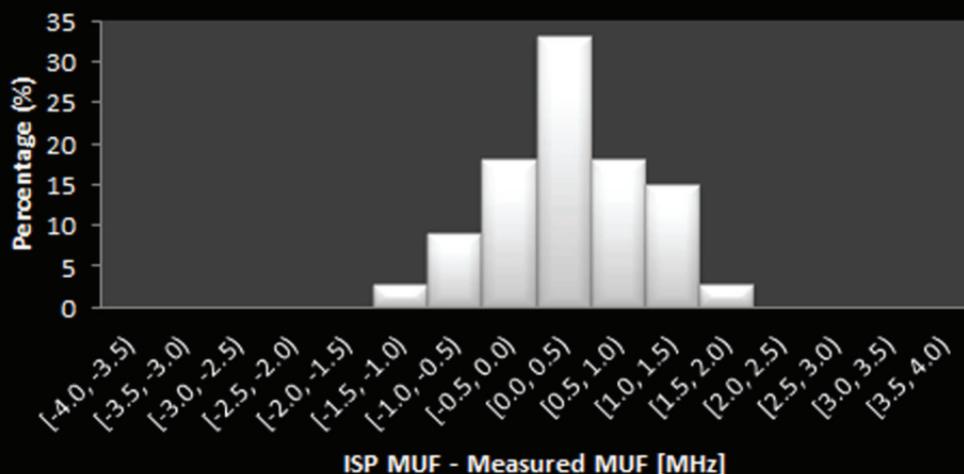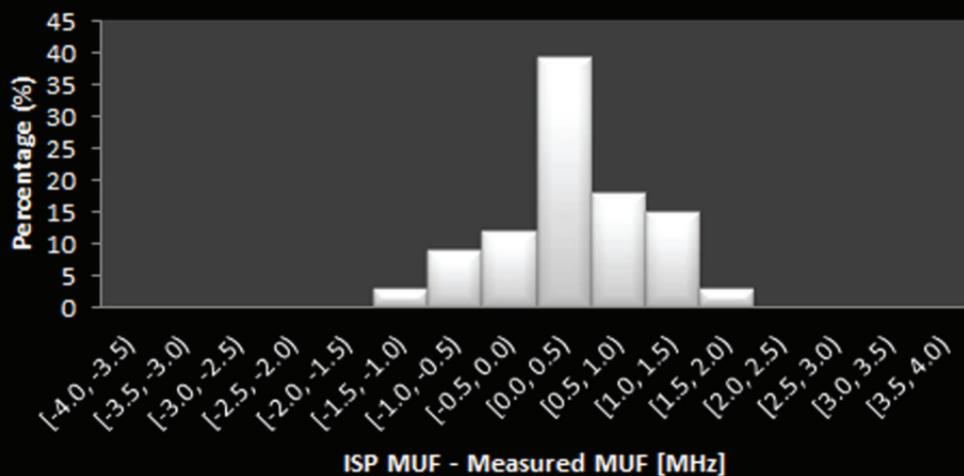

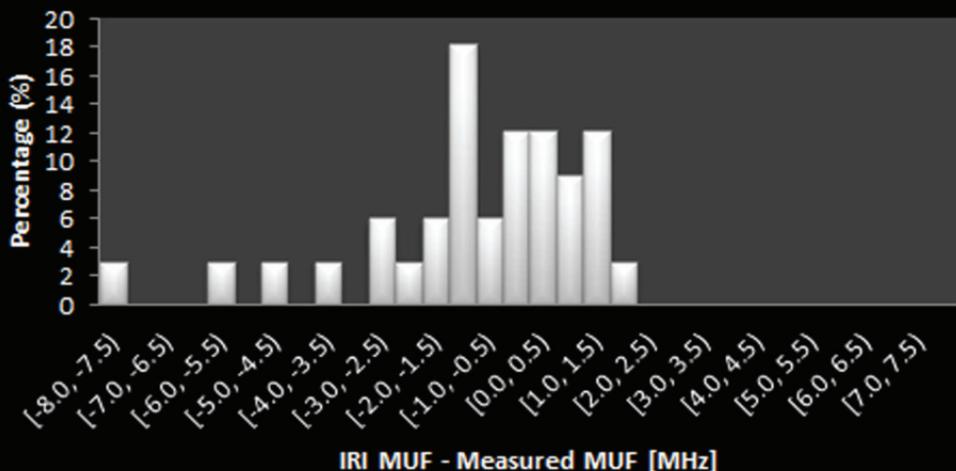
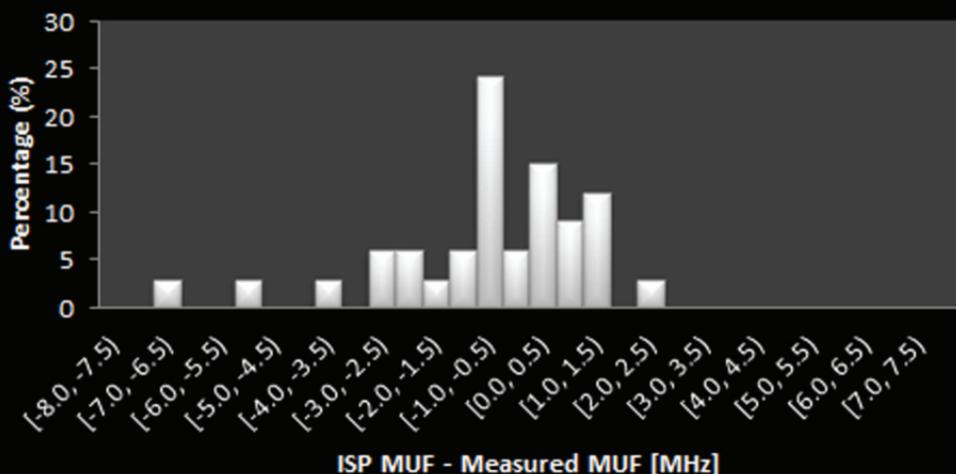
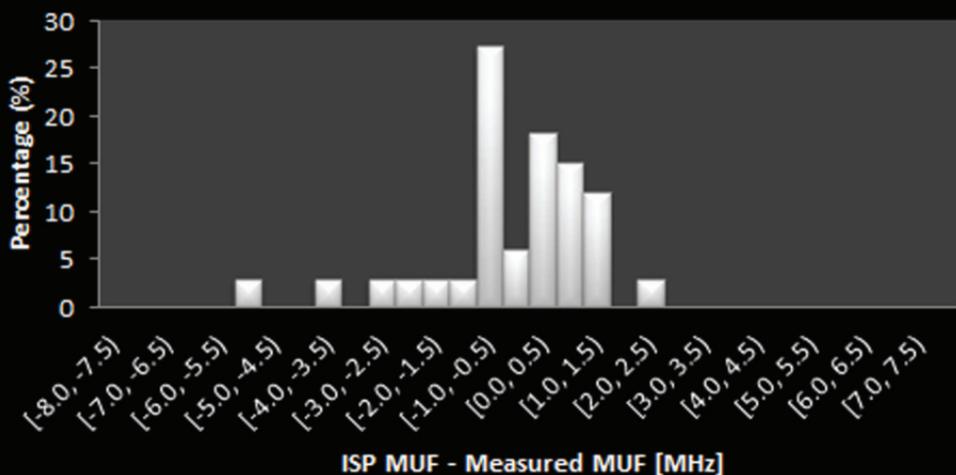

**Table 1.a – NO HORGRAD**

| DATE AND TIME [UT] | IRI MUF [MHz] | ISP MUF [MHz] (*Autoscaled Rome*) | Measured MUF [MHz] | IRI MUF-Measured MUF [MHz] | ISP MUF-Measured MUF [MHz] (*Autoscaled Rome*) |
|---|---|---|---|---|---|
| **23 JUN 2011 - 17:00** | 13.7 | 11.8 | 10.7 | 3.0 | <u>1.1</u> |
| **23 JUN 2011 - 19:00** | 12.9 | 14.9 | 14.9 | -2.0 | <u>0.0</u> |
| **23 JUN 2011 - 23:00** | 10.0 | 8.6 | 8.6 | 1.4 | <u>0.0</u> |
| **25 JUN 2011 - 10:00** | 14.6 | 14.6 | 13.4 | <u>1.2</u> | 1.2 |
| **25 JUN 2011 - 20:00** | 12.2 | 13.5 | 13.4 | -1.2 | <u>0.1</u> |
| **25 JUN 2011 - 23:00** | 9.9 | 11.7 | 11.4 | -1.5 | <u>0.3</u> |
| **26 JUN 2011 - 00:00** | 9.4 | 9.4 | 9.0 | <u>0.4</u> | 0.4 |
| **26 JUN 2011 - 01:00** | 9.0 | 8.8 | 8.8 | 0.2 | <u>0.0</u> |
| **26 JUN 2011 - 02:00** | 8.8 | 8.8 | 8.2 | <u>0.6</u> | 0.6 |
| **03 JUL 2011 - 17:00** | 13.6 | 15.2 | 15.1 | -1.5 | <u>0.1</u> |
| **04 JUL 2011 - 19:00** | 12.9 | 13.6 | 13.8 | -0.9 | <u>-0.2</u> |
| **04 JUL 2011 - 20:00** | 12.2 | 13.7 | 13.2 | -1.0 | <u>0.5</u> |
| **06 JUL 2011 - 12:00** | 14.4 | 14.6 | 13.7 | <u>0.7</u> | 0.9 |
| **06 JUL 2011 - 21:00** | 11.6 | 12.2 | 13.5 | -1.9 | <u>-1.3</u> |
| **07 JUL 2011 - 01:00** | 8.9 | 8.0 | 8.3 | 0.6 | <u>-0.3</u> |
| **07 JUL 2011 - 14:00** | 13.7 | 13.7 | 12.4 | <u>1.3</u> | 1.3 |
| **07 JUL 2011 - 15:00** | 13.5 | 13.7 | 13.0 | <u>0.5</u> | 0.7 |
| **07 JUL 2011 - 17:00** | 13.5 | 12.6 | 11.8 | 1.7 | <u>0.8</u> |
| **07 JUL 2011 - 18:00** | 13.2 | 13.6 | 13.5 | -0.3 | <u>0.1</u> |
| **07 JUL 2011 - 19:00** | 12.9 | 13.1 | 14.0 | -1.1 | <u>-0.9</u> |
| **08 JUL 2011 - 17:00** | 13.5 | 12.8 | 12.1 | 1.4 | <u>0.7</u> |
| **08 JUL 2011 - 18:00** | 13.2 | 14.8 | 14.5 | -1.3 | <u>0.3</u> |
| **08 JUL 2011 - 19:00** | 12.8 | 16.0 | 16.7 | -3.9 | <u>-0.7</u> |
| **08 OCT 2011 - 06:15** | 14.8 | 17.3 | 15.9 | <u>-1.1</u> | 1.4 |
| **08 OCT 2011 - 06:45** | 14.7 | 18.0(*18.0*) | 17.7 | -3.0 | <u>0.3 (*0.3*)</u> |
| **08 OCT 2011 - 10:30** | 19.8 | 20.3(*20.6*) | 20.5 | -0.7 | <u>-0.2(*0.1*)</u> |
| **08 OCT 2011 - 20:00** | 9.0 | 8.7(*10.6*) | 9.2 | <u>-0.2</u> | -0.5(*1.4*) |
| **08 OCT 2011 - 23:45** | 7.6 | 6.8(*6.8*) | 7.7 | <u>-0.1</u> | -0.9(*-0.9*) |
| **09 OCT 2011 - 02:00** | 7.5 | 6.5(*6.9*) | 6.9 | 0.6 | <u>-0.4(*0.0*)</u> |
| **09 OCT 2011 - 02:15** | 7.5 | 6.4(*6.4*) | 6.8 | 0.7 | <u>-0.4(*-0.4*)</u> |
| **09 OCT 2011 - 03:00** | 7.2 | 6.8(*7.0*) | 6.7 | 0.5 | <u>0.1(*0.3*)</u> |
| **09 OCT 2011 - 05:00** | 10.7 | 13.5(*11.8*) | 12.2 | -1.5 | <u>1.3(*-0.4*)</u> |
| **09 OCT 2011 - 06:30** | 14.8 | 16.9 | 15.1 | <u>-0.3</u> | 1.8 |

**Table 1.b – WITH HORGRAD**

| DATE AND TIME [UT] | IRI MUF [MHz] | ISP MUF [MHz] (*Autoscaled Rome*) | Measured MUF [MHz] | IRI MUF-Measured MUF [MHz] | ISP MUF-Measured MUF [MHz] (*Autoscaled Rome*) |
|---|---|---|---|---|---|
| **23 JUN 2011 - 17:00** | 9.3 | 11.7 | 10.7 | -1.4 | <u>1.0</u> |
| **23 JUN 2011 - 19:00** | 12.9 | 11.0 | 14.9 | <u>-2.0</u> | -3.9 |
| **23 JUN 2011 - 23:00** | 10.0 | 8.6 | 8.6 | 1.4 | <u>0.0</u> |
| **25 JUN 2011 - 10:00** | 14.8 | 14.6 | 13.4 | 1.4 | <u>1.2</u> |
| **25 JUN 2011 - 20:00** | 5.7 | 10.5 | 13.4 | -7.7 | <u>-2.9</u> |
| **25 JUN 2011 - 23:00** | 10.0 | 10.4 | 11.4 | -1.4 | <u>-1.0</u> |
| **26 JUN 2011 - 00:00** | 9.2 | 8.3 | 9.0 | <u>0.2</u> | -0.7 |
| **26 JUN 2011 - 01:00** | 8.7 | 8.2 | 8.8 | <u>-0.1</u> | -0.6 |
| **26 JUN 2011 - 02:00** | 8.6 | 8.8 | 8.2 | <u>0.4</u> | 0.6 |
| **03 JUL 2011 - 17:00** | 13.6 | 9.6 | 15.1 | <u>-1.5</u> | -5.5 |
| **04 JUL 2011 - 19:00** | 10.9 | 12.8 | 13.8 | -2.9 | <u>-1.0</u> |
| **04 JUL 2011 - 20:00** | 12.4 | 13.6 | 13.2 | -0.8 | <u>0.4</u> |
| **06 JUL 2011 - 12:00** | 14.7 | 14.0 | 13.7 | 1.0 | <u>0.3</u> |
| **06 JUL 2011 - 21:00** | 11.6 | 11.9 | 13.5 | -1.9 | <u>-1.6</u> |
| **07 JUL 2011 - 01:00** | 9.0 | 6.9 | 8.3 | <u>0.7</u> | -1.4 |
| **07 JUL 2011 - 14:00** | 13.7 | 13.6 | 12.4 | 1.3 | <u>1.2</u> |
| **07 JUL 2011 - 15:00** | 13.5 | 13.5 | 13.0 | 0.5 | <u>0.5</u> |
| **07 JUL 2011 - 17:00** | 13.5 | 9.6 | 11.8 | <u>1.7</u> | -2.2 |
| **07 JUL 2011 - 18:00** | 12.0 | 12.6 | 13.5 | -1.5 | <u>-0.9</u> |
| **07 JUL 2011 - 19:00** | 8.2 | 13.2 | 14.0 | -5.8 | <u>-0.8</u> |
| **08 JUL 2011 - 17:00** | 9.7 | 12.6 | 12.1 | -2.4 | <u>0.5</u> |
| **08 JUL 2011 - 18:00** | 13.0 | 14.8 | 14.5 | -1.5 | <u>0.3</u> |
| **08 JUL 2011 - 19:00** | 12.9 | 16.0 | 16.7 | -3.8 | <u>-0.7</u> |
| **08 OCT 2011 - 06:15** | 14.8 | 17.1 | 15.9 | <u>-1.1</u> | 1.2 |
| **08 OCT 2011 - 06:45** | 14.8 | 10.8 (*18.1*) | 17.7 | -2.9 | -6.9 (<u>*0.4*</u>) |
| **08 OCT 2011 - 10:30** | 19.8 | 17.8 (*21.0*) | 20.5 | -0.7 | -2.7 (<u>*0.5*</u>) |
| **08 OCT 2011 - 20:00** | 9.0 | 7.1 (*9.7*) | 9.2 | <u>-0.2</u> | -2.1 (*0.5*) |
| **08 OCT 2011 - 23:45** | 7.6 | 6.8 (*6.9*) | 7.7 | <u>-0.1</u> | -0.9 (-*0.8*) |
| **09 OCT 2011 - 02:00** | 7.1 | 6.5 (*6.1*) | 6.9 | <u>0.2</u> | -0.4 (-*0.8*) |
| **09 OCT 2011 - 02:15** | 7.5 | 6.4 (*6.3*) | 6.8 | 0.7 | <u>-0.4 (-*0.5*)</u> |
| **09 OCT 2011 - 03:00** | 6.7 | 6.8 (*7.0*) | 6.7 | <u>0.0</u> | 0.1 (*0.3*) |
| **09 OCT 2011 - 05:00** | 7.5 | 11.1 (*11.7*) | 12.2 | -4.7 | <u>-1.1 (-*0.5*)</u> |
| **09 OCT 2011 - 06:30** | 14.9 | 17.1 | 15.1 | <u>-0.2</u> | 2.0 |